\documentclass[
aps,%
final,%
prd,%
twocolumn,%
showpacs,%
superscriptaddress]%
{revtex4-1}

\usepackage{amsmath}
\usepackage{amssymb}
\usepackage{graphicx}

\usepackage{subfigure}
\usepackage{sidecap}

\begin{document}

\bibliographystyle{ieeetr}

\title{Neutrino spectra evolution during proto-neutron star deleptonization}

\author{T.~Fischer}
\email{t.fischer@gsi.de}
\affiliation{GSI Helmholtzzentrum f\"ur Schwerioneneforschung, Planckstra{\ss}e~1, 64291 Darmstadt, Germany} 
\affiliation{Institut f{\"u}r Kernphysik, Technische Universit{\"a}t Darmstadt, Schlossgartenstra{\ss}e 9, 64289 Darmstadt, Germany}

\author{G.~Mart{\'i}nez-Pinedo}
\affiliation{Institut f{\"u}r Kernphysik, Technische Universit{\"a}t Darmstadt, Schlossgartenstra{\ss}e 9, 64289 Darmstadt, Germany}
\affiliation{GSI Helmholtzzentrum f\"ur Schwerioneneforschung, Planckstra{\ss}e~1, 64291 Darmstadt, Germany} 

\author{M.~Hempel}
\affiliation{Department of Physics, University of Basel, Klingelbergstra{\ss}e 82, 4056 Basel, Switzerland}  

\author{M.~Liebend\"orfer}
\affiliation{Department of Physics, University of Basel, Klingelbergstra{\ss}e 82, 4056 Basel, Switzerland}  

\begin{abstract}
The neutrino-driven wind, which occurs after the onset of a core-collapse supernova explosion, has long been considered as the possible site for the synthesis of heavy $r$-process elements in the Universe. Only recently, it has been possible to simulate supernova explosions up to $\sim10$~seconds, based on three-flavor Boltzmann neutrino transport. These simulations show that the neutrino luminosities and spectra of all flavors are very similar and their difference even decreases during the deleptonization of the proto-neutron star. As a consequence, the ejecta are always proton rich which rules out the possible production of heavy $r$-process elements ($Z>56$).  We perform a detailed analysis of the different weak processes that determine the neutrino spectra. Non-electron flavor (anti)neutrinos are produced and interact only via neutral-current processes, while electron (anti)neutrinos have additional contributions from charge-current processes. The latter are dominated by $\nu_e$-absorption on neutrons and $\bar\nu_e$-absorption on protons. At early times, charge-current processes are responsible for spectral differences between $\nu_e$, $\bar\nu_e$ and $\nu_{\mu/\tau}$. However, as the region of neutrino decoupling moves to higher densities during deleptonization, charge-current reactions are suppressed by final state Pauli-blocking. $\bar\nu_e$ absorption on protons is suppressed due to the continuously increasing chemical potential of the neutrons. $\nu_e$ absorption on neutrons is blocked by the increasing degeneracy of the electrons. These effects result in negligible contributions from charge-current reactions on timescales on the order of tens of seconds, depending on the progenitor star. Hence, the neutrino spectra are mainly determined from neutral-current processes which do not distinguish between the different flavors and results in the convergence of the spectra. These findings are independent of the charge-current reaction rates used. It rules out the possibility of neutron-rich ejecta at late times and the production of heavy $r$-process elements from non-rotating and not magnetized proto-neutron stars.
\end{abstract}

\date{\today}


\maketitle

\section{Introduction}

Core collapse supernova explosions of stars more massive than $8$~M$_\odot$ is an active subject of research in theoretical astrophysics~\cite{Janka:2007}. They are related to the revival of the stalled shock, which forms when the collapsing core bounces back at nuclear densities. Right after the formation, the shock looses energy due to the dissociation of heavy nuclei and to the release of the $\nu_e$-deleptonization burst produced by electron captures on protons. At this moment the shock becomes an accretion front and stalls at a radius between 100 and 200 km. Different mechanism have been suggested to revive the shock and finally explode the star. These are the magneto-rotational mechanism~\cite{LeBlanc:1970kg,Moiseenko:2007zz,Takiwaki:2007sf}, the acoustic mechanism due to the dumping of sound waves~\cite{Burrows:2005dv} and the delayed neutrino-heating mechanism~\cite{Bethe:1985ux}.  Recently, it has been shown that a phase transition to deconfined quark matter can lead to an additional collapse during which a second hydrodynamic shock wave forms that triggers the explosion~\cite{Sagert:2008ka,Fischer:2011}.

Independently of the explosion mechanism, a proto-neutron star (PNS) forms at the supernova core immediately after core bounce. The initially hot and lepton-rich PNS cools via continuous emission of neutrinos of all flavors in a period of several tens of seconds after the explosion has been launched. Neutrino-absorption processes at the PNS surface deposit energy, which drives a matter outflow known as the neutrino-driven wind. It has been investigated as a possible site for the production of heavy $r$-process elements~\cite{Woosley:1994ux,Takahashi:1994yz}. These pioneering works where followed by analytic~\cite{Qian:1996xt}, parametric~\cite{Hoffman:1996aj} and steady state wind models~\cite{Otsuki:1999kb,Thompson:2001ys}. They showed that the production of heavy $r$-process elements requires a mass outflow on a short dynamical time scale (a few milliseconds), high entropies per baryon (above 150~k$_\text{B}$) and a low proton-to-baryon ratio, i.e. electron fraction $Y_e < 0.5$. Recent hydrodynamics simulations~\cite{Arcones:2006uq,Arcones:2011a} showed that short dynamical timescales can in fact be achieved but fail however to obtain the necessary entropies at times relevant for $r$-process nucleosynthesis~\cite{Kuroda.Wanajo.Nomoto:2008}.

The nucleosynthesis outcome is very sensitive to the electron fraction of the ejected matter~\cite{Arcones:2011b} that is determined by the competition between electron neutrino absorption on neutrons and antineutrino absorption on protons and their inverse reactions. Deep in the PNS interior, neutrinos are in chemical equilibrium with matter. However, at the PNS surface where temperature and density drop, neutrinos decouple from matter. Since $\mu$ and $\tau$ neutrinos interact only via neutral-current reactions, they are the first to decouple at the highest temperature. For the very neutron-rich conditions found at the PNS surface, $\bar\nu_e$ decouple before $\nu_e$. The neutrino spectra reflect local properties of matter at the position where they decouple. Hence, one expects the following hierarchy of neutrino energies: $\varepsilon_{\nu_{\mu,\tau}} > \varepsilon_{\bar{\nu}_e} > \varepsilon_{\nu_e}$~\cite{Keil:2003}, where $\varepsilon = \langle E^2\rangle/\langle E \rangle$. $\langle E \rangle$ is the mean neutrino energy and $\langle E^2 \rangle$ is the square value of the root-mean-square (rms) energy.. The fact that $\bar\nu_e$ have larger average energies than $\nu_e$ suggest that the ejecta will be neutron rich. However, one also has to consider the other energy scale in the problem, i.e. the neutron-to-proton mass difference, $\triangle=m_n-m_p=1.2935$~MeV. It turns out that neutron-rich ejecta are only obtained when $\varepsilon_{\bar{\nu}_e} - \varepsilon_{\nu_e} > 4\,\triangle$~\cite{Qian:1996xt,Frohlich:2006} in the presence of similar electron neutrino and antineutrino luminosities.

The treatment of neutrino transport and neutrino matter interactions improved over the years significantly. Simulations of core-collapse supernovae, that include three-flavor Boltzmann neutrino transport, showed rather small neutrino energy differences between $\bar\nu_e$ and $\nu_e$ and hence generally proton-rich conditions were obtained during the early evolution after the onset of explosion~\cite{Liebendoerfer:2001a,Buras:2005rp}. More recently, it has been possible to simulate core-collapse supernovae consistently through all phases up to 20~seconds after the onset of explosion~\cite{Fischer:2009af,Huedepohl:2010}, using general relativistic radiation hydrodynamics that employs three-flavor Boltzmann neutrino transport and a sophisticated equation of state. It has been shown that the energy difference between $\bar\nu_e$ and $\nu_e$ decreases continuously after the explosion has been launched and hence generally proton-rich conditions remain for more than 10~seconds. This rules out initial expectations that the ejecta can become neutron-rich at late times. It leaves the $\nu p$-process~\cite{Frohlich:2005ys,Pruet:2006,Wanajo:2006ec} as the only nucleosynthesis process that can occur in neutrino-driven winds from non-rotating and not magnetized PNSs.

In this article, we analyze the neutrino spectra obtained in the current simulations focussing in understanding the reasons behind the reduction of the energy difference between $\bar\nu_e$ and $\nu_e$ during the deleptonization of the PNS. We perform a detailed analysis of the main reactions determining the opacities or inverse mean free paths for the different neutrino flavors. Special attention is devoted to neutrino decoupling from matter, in particular the determination of the position of the neutrinospheres which evolves during the PNS deleptonization to continuously higher densities. We also present an alternative analysis, based on the formalism introduced in ref.~\cite{Liebendoerfer:2004}, that determines the contributions to the neutrino luminosity from different processes and regions of the PNS. Both approaches show that at late times during the deleptonization, the spectra of all neutrino species forms at very similar radii and is determined by neutral-current processes that are insensitive to the neutrino flavor.

The paper is organizes as follows. In section~II we will introduce the procedure used to calculate the inverse mean free paths for the relevant weak interaction processes considered, based on the neutrino distribution functions obtained in three-flavor Boltzmann neutrino transport.  We will furthermore discuss neutrino decoupling and introduce the concept of a scattering atmosphere following the work of ref.~\cite{Raffelt:2001}.  In section~III, we will apply this analysis and discuss the evolutionary behavior of the neutrino spectra obtained in core collapse supernova simulations of the low-mass 8.8~M$_\odot$ O-Ne-Mg-core~\cite{Fischer:2009af}.  In section~IV we extend the analysis from section~III to the 18~M$_\odot$ iron-core progenitor from ref~\cite{Fischer:2009af} and find qualitatively similar results. In section~V is show the evolution of the neutrino spectra at different times during the deleptonization. While we explored opacities in section~III and IV, in section~VI we analyze the neutrino emission, focusing on the location at which different contributions to the luminosity at infinity appear. We close the manuscript with a summary in section~VII.
%

\section{Neutrino spectra}

We analyze data obtained in the core collapse supernova simulations of massive stars published in ref.~\cite{Fischer:2009af}. Stars in the mass range of 8.8--18~M$_\odot$ were evolved consistently through core collapse, bounce and explosion including the neutrino-driven wind phase for more than 20~seconds after core bounce.  It includes most part of the PNS deleptonization, i.e. cooling via the emission of neutrinos of all flavors.  For these models, the baryonic equation of state from Shen~\emph{et~al.} was used~\cite{Shen:1998gg}. It is based on relativistic mean field theory and the Thomas-Fermi approximation for heavy nuclei.  Contributions from ($e^-,\,e^+$) and photons where added following ref.~\cite{Timmes:1999}.

The simulations were based on general relativistic radiation hydrodynamics and three-flavor Boltzmann neutrino transport in spherical symmetry~\cite{Mezzacappa:1993gn,Mezzacappa:1993gm,Mezzacappa:1993gx,Liebendoerfer:2004}. The model is based on the following line element,
\begin{equation}
ds^2 = -\alpha^2 dt^2 + \left(\frac{r'}{\Gamma}\right)^2 da^2 + d\Omega,
\end{equation}
which describes non-stationary and spherically symmetric spacetime, in coordinates $t$ (system time) and $a$ (enclosed baryon mass), where $d\Omega=r^2 ( d\theta + \sin^2\theta d\phi)$ describes a 2-sphere of radius $r(t,a)$. $\alpha(t,a)$ and $\Gamma(t,a)$ are the metric functions and $r(t,a)'=\partial r/\partial a$. With the following stress-energy tensor,
\begin{subequations}
\begin{eqnarray}
T^{tt} &=& \rho(1+e+J), \\
T^{at} = T^{ta} &=& \rho H, \\
T^{aa} &=& p + \rho K, \\
T^{\theta\theta} = T^{\phi\phi} &=& p + \frac{1}{2}\rho(J-K),
\end{eqnarray}
\end{subequations}
with rest-mass density $\rho$, internal energy density $e$ and matter pressure $p$, the equations for energy- and momentum-conservation are obtained via the covariant derivative of the stress-energy tensor as follows, $\nabla_k \ T^{kl}=0$. The quantities $J$, $H$ and $K$ are the zero, first and second angular moments of the neutrino distribution functions, $f_\nu(t,a,\mu,E)$, for more details see~\cite{Liebendoerfer:2004}. The mater velocity is given by the following relation, $u=\partial r / \alpha\partial t$.

Neutrino transport determines the evolution of these neutrino distribution functions. In spherical symmetry, their momentum-dependency is given by the cosine of the neutrino phase-space propagation angle ($\mu$) and the neutrino energy ($E$), for each neutrino flavor $\nu\in\{\nu_e,\bar{\nu}_e,\nu_{\mu/\tau},\bar{\nu}_{\mu/\tau}\}$. Their evolution is determined by solving the Boltzmann transport equation, assuming ultra-relativistic massless fermions, in a co-moving reference frame. Defining the specific distribution functions, $F_\nu=f_\nu/\rho$, the Boltzmann transport equation can be expressed as follows~\cite{Lindquist:1966,Liebendoerfer:2004} 
\begin{subequations}
\begin{equation}
\frac{\partial F(\mu,E)}{\alpha\partial t} =
\frac{\mu}{\alpha} \frac{\partial}{\partial a}
\left ( 4\pi r^{2}\alpha\rho \, F(\mu,E)\right )
\label{eq:Boltz01}
\end{equation}
\begin{equation}
+
\Gamma
\left (
\frac{1}{r}-\frac{1}{\alpha}\frac{\partial\alpha}{\partial r}
\right )
\frac{\partial}{\partial\mu}
\left [
\left ( 1-\mu^{2} \right ) F(\mu,E)
\right ]
\label{eq:Boltz02}
\end{equation}
\begin{equation}
+
\left (
\frac{\partial\ln\rho}{\alpha\partial t} + \frac{3u}{r}
\right )
\frac{\partial}{\partial\mu}
\left [
\mu \left ( 1-\mu^{2} \right ) F(\mu,E)
\right ]
\label{eq:Boltz03}
\end{equation}
\begin{equation}
-
\mu\Gamma\frac{1}{\alpha}\frac{\partial\alpha}{\partial r}
\frac{1}{E^{2}}\frac{\partial}{\partial E} \left ( E^{3} F(\mu,E) \right )
\label{eq:Boltz04}
\end{equation}
\begin{equation}
+
\left [
\mu^{2}
\left (
\frac{\partial \ln\rho}{\alpha\partial t} + \frac{3u}{r}
\right )
-\frac{u}{r}
\right ]
\frac{1}{E^{2}}\frac{\partial}{\partial E} \left ( E^{3} F(\mu,E) \right )
\label{eq:Boltz05}
\end{equation}
\begin{equation}
+ \left ( \frac{dF(\mu,E)}{\alpha dt} \right )_{\text{collision}}.
\label{eq:Boltzcol}
\end{equation}
\label{eq:Boltz}
\end{subequations}
Note that the neutrino distribution functions depend also on the system time $t$ and radial coordinate (or enclosed baryon mass $a$), which are not shown here for simplicity. Expressions~(\ref{eq:Boltz01})--(\ref{eq:Boltz05}) are the transport terms of the Boltzmann equation, while expression~(\ref{eq:Boltzcol}) is the collisional term. The latter is given by weak processes that change the neutrino distributions, as follows
\begin{widetext}
\begin{subequations}
\begin{equation}
\left ( \frac{dF_\nu(\mu,E)}{\alpha dt} \right )_{\text{collision}}
=
j(E)\left(\frac{1}{\rho}-F_\nu(\mu,E)\right) - \frac{c}{\lambda(E)}F_\nu(\mu,E)
\label{eq:Boltzcol01}
\end{equation}
\begin{eqnarray}
&+&
  \frac{E^{2}}{c(hc)^{3}} \int d\mu' \int d\phi\,\,R_{\text{IS}, \nu N/A}^0(E,E,\cos\theta)\,F_\nu(\mu',E)
 - \frac{E^{2} F_\nu(\mu,E)}{c(hc)^{3}}  \int d\mu'\int d\phi\,\,R_{\text{IS}, \nu N/A}^0(E,E,\cos\theta)
\label{eq:Boltzcol02}
\\
&+& \left(\frac{1}{\rho}-F_\nu(\mu,E)\right)\frac{1}{c(hc)^{3}}
  \int  E'^2\,dE' \int  d\mu' \int d\phi \, R_{\nu e^\pm}^\text{in}(E,E',\cos\theta)\,F_\nu(\mu',E')
  \nonumber
  \\
&-& F_\nu(\mu,E) \frac{1}{c(hc)^{3}}
  \int  E'^2\,dE' \int  d\mu' \int d\phi \, R_{\nu e^\pm}^\text{out}(E,E',\cos\theta)\,
  \left(\frac{1}{\rho}-F_\nu(\mu',E')\right)
  \label{eq:Boltzcol03}
\\
&+& \left(\frac{1}{\rho}-F_\nu(\mu,E)\right) \frac{1}{c(hc)^{3}}
  \int  E'^2\,dE' \int  d\mu' \int d\phi \,\,
  R_{\nu\bar\nu}^\text{p}(E,E',\cos\theta) \left(\frac{1}{\rho}-F_{\bar\nu}(\mu',E')\right)
  \nonumber
  \\
&-& F_\nu(\mu,E) \frac{1}{c(hc)^{3}}
  \int  E'^2\,dE' \int  d\mu' \int d\phi \,\,R_{\nu\bar\nu}^\text{a}(E,E',\cos\theta) F_{\bar\nu}(\mu',E').
  \label{eq:Boltzcol04}
\end{eqnarray}
\end{subequations}
\end{widetext}
In the Boltzmann transport representation, weak processes are characterized by reaction rates. The relevant processes are listed in Table~\ref{table-nu-reactions}. For charge-current reactions these are the {\em emissivity}, $j(E)$ and the inverse mean free path, $1/\lambda(E)$. For neutral current processes we employ {\em scattering kernels}, $R(E,E',\cos\theta)$. They define the probability of a neutrino of energy $E$ and momentum space coordinates $(\mu,\varphi)$ being scattered to an energy $E'$ and coordinates $(\mu',\varphi')$. Apart of the energies the scattering probability is a function of the scattering angle~\cite{Bruenn:1985en},
\begin{equation}
  \label{eq:scatangle}
  \cos\theta = \mu\mu' + \sqrt{(1-\mu^2)(1-\mu'^2)} \cos\phi, \quad \phi\equiv \varphi - \varphi'.
\end{equation}
The elastic processes considered are neutrino scattering on nucleons ($N$) and nuclei ($A$), $R_{\text{IS}, \nu N/A}(E,E',\cos\theta)$. The inelastic processes are neutrino scattering on electrons/positrons, $R_{\nu e^\pm}^\text{in/out}(E,E',\cos\theta)$, for in-scattering and out-scattering, and pair processes, $R_{\nu\bar\nu}^\text{a/p}(E,E',\cos\theta)$, for the absorption of a neutrino pair (a) and for the production of a neutrino pair (p). The reaction rates for all processes will be further discussed below in section~II A.

\begin{table*}[htp]
\centering
\caption{Neutrino reactions considered, including reactions rates and references.}
\begin{tabular}{cccc}
\hline
\hline
&
Reaction\footnote{Note:
  $\nu=\{\nu_e,\bar{\nu}_e,\nu_{\mu/\tau},\bar{\nu}_{\mu/\tau}\}$ 
and $N=\{n,p\}$}
&
Mean free path/Reaction kernel \ \
&
References
\\
\hline
1
&
$\nu_e + n \rightarrow p + e^-$
&
$1/\lambda_{\nu_e n}(E)$
&
\cite{Bruenn:1985en}\\
2
&
$\bar{\nu}_e + p \rightarrow n + e^+$
&
$1/\lambda_{\bar{\nu}_e p}(E)$
&
\cite{Bruenn:1985en}
\\
3
&
$\nu_e + (A,Z-1) \rightarrow (A,Z) + e^-$
&
$1/\lambda_{\nu_e A}(E)$
&
\cite{Bruenn:1985en}
\\
4
&
$\nu + N \rightarrow \nu' + N$
&
$R_{\text{IS}, \nu N}^0(E,E,\cos\theta)$
&
\cite{Bruenn:1985en}
\\
5
&
$\nu + (A,Z) \rightarrow \nu' + (A,Z)$
&
$R_{\text{IS}, \nu A}^0(E,E,\cos\theta)$
&
\cite{Bruenn:1985en}
\\
6
&
$\nu + e^\pm \rightarrow \nu' + e^\pm$
&
$R_{\nu e^{\pm}}^\text{in}(E,E',\cos\theta)$
&
\cite{Bruenn:1985en,Mezzacappa:1993gx,Mezzacappa:1993gm}
\\
7
&
$\nu + \bar{\nu} \rightarrow e^- + e^+$
&
$R_{\nu\bar{\nu}}^\text{a}(E,E',\cos\theta)$
&
\cite{Bruenn:1985en,Mezzacappa:1999}
\\
8
&
$\nu + \bar{\nu} + N + N \rightarrow N + N$
&
$R_{\nu\bar{\nu}NN}^\text{a}(E,E',\cos\theta)$
&
\cite{Hannestad:1997gc}
\\
\hline
\end{tabular}
\label{table-nu-reactions}
\end{table*}

\subsection{Reaction rates and opacities}

Let us first focus on charge-current reactions and discuss neutral current reactions further below. Assuming a composition given by free protons and neutrons, the change in the net electron fraction, $Y_e = Y_{e^-}- Y_{e^+}$ is given by 
\begin{equation}
  \label{eq:dotye}
  \dot{Y}_e = -\left( r_{e^- p} + r_{\bar{\nu}_e p} \right) Y_p + \left( r_{e^+ n} + r_{\nu_e n} \right) Y_n.  
\end{equation}
where $Y_p$ and $Y_n$ are the proton and neutron abundances and the $r$'s are rates for electron capture ($e^- p$), antineutrino absorption ($\bar{\nu}_e p$), positron capture ($e^+ n$) and neutrino absorption ($\nu_e n$). 

Boltzmann neutrino transport uses energy dependent neutrino emissivities, $j(E)$, and opacities, $\chi(E)$ (see equation~(\ref{eq:Boltzcol01})). The latter is equal to the inverse mean free path, $1/\lambda(E)$. They depend on the distribution functions for $e^\pm$ and nucleons that are assumed to be in thermal equilibrium. The rates which appear in equation~\eqref{eq:dotye} are then expressed as follow
\begin{subequations}
  \label{eq:rates}
  \begin{equation}
    \label{eq:rec}
    r_{e^- p} = \frac{2\pi}{(h c)^3 n_p} \int d\mu\,dE\,E^2 j_{\nu_e}(E)
    [1-f_{\nu_e}(\mu,E)],  
  \end{equation}
  \begin{equation}
    \label{eq:rnubarp}
    r_{\bar{\nu}_e p} = \frac{2\pi}{(hc)^3 n_p} \int d\mu\,dE\,E^2
    c \frac{f_{\bar{\nu}_e}(\mu,E)}{\lambda_{\bar{\nu}_e}(E)},
  \end{equation}
  \begin{equation}
    \label{eq:rpc}
    r_{e^+ n} = \frac{2\pi}{(h c)^3 n_n} \int d\mu\,dE\,E^2 j_{\bar{\nu}_e}(E)
    [1-f_{\bar{\nu}_e}(\mu,E)],  
  \end{equation}
  \begin{equation}
    \label{eq:rnun}
    r_{\nu_e n} = \frac{2\pi}{(hc)^3 n_n} \int d\mu\,dE\,E^2
    c \frac{f_{\nu_e}(\mu,E)}{\lambda_{\nu_e}(E)}, 
  \end{equation}
\end{subequations}
where $n_p$ and $n_n$ are the proton and neutron number densities. For the neutrino and antineutrino emissivities and mean-free paths we use the analytical expressions given in appendix C of ref.~\cite{Bruenn:1985en}. Note furthermore, the emissivity and opacity (or equivalent the mean-free path) are related by detailed balance as follows~\cite{Bruenn:1985en},
\begin{subequations}
\label{eq:deba}
\begin{equation}
\label{eq:debanu}
j_{\nu_e}(E) = \exp\left\{-\frac{E-(\mu_e-\mu_Q)}{k T}\right\} \frac{c}{\lambda_{\nu_e}(E)},
\end{equation}
\begin{equation}
j_{\bar\nu_e}(E) = \exp\left\{-\frac{E+(\mu_e-\mu_Q)}{k T}\right\} \frac{c}{\lambda_{\bar\nu_ep}(E)},
\end{equation}
\end{subequations}
where $\mu_Q = \mu_n - \mu_p$. Note that we also use the reaction rates for charge-current processes with heavy nuclei from ref.~\cite{Bruenn:1985en}, which are based on a simplified description of the Gamow-Teller transition assuming an average nucleus with mean charge and mass. This has been improved recently, taking into account the distribution of heavy nuclei in nuclear statistical equilibrium~\cite{Langanke:2003,Hix:2003}.

In a similar fashion as for the charge-current reactions, we can define mean free paths for the scattering processes:
\begin{eqnarray}
  \label{eq:mscattering}
  \frac{1}{\lambda(E,\mu)} & = & \frac{1}{c (hc)^3} \int_0^\infty dE' E'^2
  \int_{-1}^1 d\mu' [1-f(\mu',E')] \nonumber \\ 
       & & \times \int_0^{2\pi} d\phi \,\, R^\text{in}(E,E',\cos\theta). 
\end{eqnarray}
They are given in terms of the reaction kernels for the different reactions, $R(E,E',\cos \theta)$ (see expressions~(\ref{eq:Boltzcol02})--(\ref{eq:Boltzcol04})). The scattering processes included in our simulations are scattering on nucleons and nuclei, electrons and positrons (reactions 4, 5 and 6 in table~\ref{table-nu-reactions}). Scattering on nucleons and nuclei is treated in the elastic approximation also known as {\em isoenergetic scattering} (IS). In this case the scattering kernel reduces to $R_{\text{IS}}(E,E',\cos\theta) = R^0_{\text{IS}}(E,E,\cos\theta)\delta(E-E')$ and equation~\eqref{eq:mscattering} gives:
\begin{eqnarray}
  \label{eq:isosc}
  \frac{1}{\lambda_{\text{IS}}(E,\mu)} &=& \frac{1}{c (hc)^3} E^2
  \int_{-1}^{1} d\mu' \int_0^{2\pi} d\phi
  \nonumber
  \\
  && \times R^0_{\text{IS}}(E,E,\cos\theta) [1-f(\mu',E)].
\end{eqnarray}
The expression we use for $R^0_{\text{IS}}(E,E,\cos\theta)$ is given in refs.~\cite{Bruenn:1985en,Mezzacappa:1993gm}. For neutrino scattering on electrons and positrons (expression 6 in table~\ref{table-nu-reactions}) we use the scattering kernels of references~\cite{Bruenn:1985en,Mezzacappa:1993gm,Mezzacappa:1993gx}.  Note that in addition to having different kernels for $\nu_e$ and $\bar\nu_e$, we also include differences between $\nu_{\mu/\tau}$ and $\bar\nu_{\mu/\tau}$.

For pair processes (expressions 7 and 8 in table~\ref{table-nu-reactions}) the relation between the mean free path and the scattering kernel as follows
\begin{eqnarray}
  \label{eq:mpair}
  \frac{1}{\lambda_\nu(E,\mu)} &=& \frac{1}{c (hc)^3} \int_0^\infty dE' E'^2
  \int_{-1}^1 d\mu' f_{\bar\nu}(\mu',E') 
  \nonumber \\
  && \times \int_0^{2\pi}d\phi R_{\nu\bar\nu}^a(E,E',\cos\theta), 
\end{eqnarray}
and similarly for antineutrinos. Table~\ref{table-nu-reactions} gives the references from where the expressions for the different scattering kernels have been obtained.  Similar to equation~\ref{eq:deba}, the reaction kernels for in- and out-scattering as well as pair production (p) and pair absorption (a) are related via detailed balance~\cite{Bruenn:1985en} as follows
\begin{eqnarray*}
R_{\nu e^\pm}^\text{in}(E,E',\cos\theta) &=& \exp\left\{\frac{-(E-E')}{kT}\right\}R_{\nu e^\pm}^\text{out}(E,E',\cos\theta), \\
R_{\nu\bar\nu}^\text{a}(E,E',\cos\theta) &=& \exp\left\{\frac{E+E'}{kT}\right\}R_{\nu\bar\nu}^\text{p}(E,E',\cos\theta).
\end{eqnarray*}

\subsection{Neutrino decoupling}

The neutrinosphere radii $R_\nu(E)$ at which neutrinos of energy $E$ and flavor $\nu$ decouple from matter, i.e. the surface of last scattering, is determined by the condition that the optical depth becomes $2/3$. The optical depth, $\tau$, is defined as follows, 
\begin{equation}
\tau_\nu(E) = \int^\infty_{R_\nu(E)} dr \frac{1}{\lambda_\nu(E)},
\label{eq:tau}
\end{equation}
integrating the total inverse mean free path, $1/\lambda_\nu(E)=\sum_i 1/\lambda_{i,\nu}(E)$, from the stellar surface towards the center, where $1/\lambda_{i,\nu}$ refers to an individual weak process $i$. Following refs~\cite{ShapiroTeukolsky:1983,Keil:2003} we use the effective neutrino opacity for energy exchange or thermalization and define:
\begin{equation}
  \frac{1}{\lambda_{\nu,\text{eff}}(E)} = \sqrt{\frac{1}{\lambda_{\nu,\text{abs}}(E)}
    \left(\frac{1}{\lambda_{\nu,\text{abs}}(E)} +
      \frac{1}{\lambda_{\nu,\text{scatt}}(E)}\right)}.
\end{equation}
The inverse absorption mean free path, $1/\lambda_{\nu,\text{abs}}(E)$, is obtained by summing contributions of processes in which neutrinos exchange energy with the medium (processes 1-3 and 6-8 in table~\ref{table-nu-reactions}). The inverse scattering mean free path contains processes where momentum of the neutrino is changed but not their energy (processes 4 and 5 in table~\ref{table-nu-reactions}).

\begin{figure}[htp]
\centering
\includegraphics[width=\columnwidth]{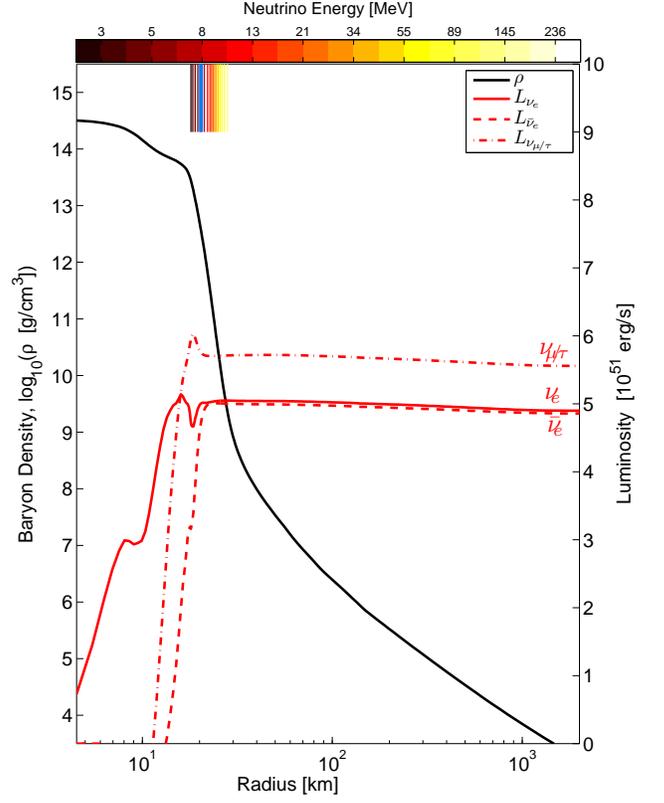}
\caption{The region of neutrino decoupling is shown via the energy-dependent transport spheres (for $\nu_e$), where the color-coding is according to the neutrino energies from $E_\text{min}=3$~MeV (black) at about 16.17~km up to $E_\text{max}=300$~MeV (white) at 49.11~km.  The energy-averaged $\nu_e$-sphere is shown in the vertical blue solid line. The radial profiles of density (solid black line) and luminosities ($\nu_e$: solid red line, $\bar\nu_e$: dashed red line, $\nu_{\mu/\tau}$: dash-dotted line) are also shown.}
\label{fig:decoupling}
\end{figure}

For those cases, scattering and pair processes, in which the opacity depends on both the energy and angle of the neutrino, we define the energy-dependent inverse mean free path as follows
\begin{equation}
  \label{eq:meanE}
  \frac{1}{\lambda_\nu(E)} = \frac{1}{n_\nu(E)} \int d\mu\,f_\nu(E,\mu)
  \frac{1}{\lambda_\nu(E,\mu)},  
\end{equation}
using the local neutrino distribution function for each neutrino flavor $\nu$ and
\begin{equation}
  \label{eq:n_nuE}
  n_\nu(E) = \int d\mu f_\nu(E,\mu).
\end{equation}

In order to determine an average neutrinosphere radius, the so-called energy sphere~\cite{Raffelt:2001}, we define an energy averaged inverse mean free path using the local neutrino phase-space distribution functions:
\begin{equation}
\left\langle\frac{1}{\lambda_{\nu}}\right\rangle
=
\frac{1}{n_\nu}
\frac{2\pi}{(h\,c)^3}
\int d\mu\,E^2 dE \,f_\nu(\mu,E) \frac{1}{\lambda_{\nu}(E,\mu)},
\end{equation}
for each neutrino flavor $\nu$ and
\begin{equation}
n_\nu = \frac{2\pi}{(h\,c)^3} \int d\mu\,E^2 dE \, f_\nu(\mu,E).
\label{eq-Nnu}
\end{equation}
is the local number density of neutrinos. The energy averaged effective mean free path is defined as
\begin{equation}
  \label{eq:lambdamuE}
  \left\langle \frac{1}{\lambda_{\nu,\text{eff}}}\right\rangle =
  \sqrt{\left\langle\frac{1}{\lambda_{\nu,\text{abs}}}\right\rangle  
    \left(\left\langle\frac{1}{\lambda_{\nu,\text{abs}}}\right\rangle +
      \left\langle\frac{1}{\lambda_{\nu,\text{scatt}}}\right\rangle\right)}. 
\end{equation}
Finally, the energy sphere radius $R_\text{ES}$ is obtained from the condition:
\begin{equation}
  \label{eq:tautherm}
  \tau{_{\text{therm}}}  (R_{\text{ES}}) = \int_{R_{\text{ES}}}^\infty dr \left\langle
    \frac{1}{\lambda_{\nu,\text{eff}}}\right\rangle \equiv \frac{2}{3} 
\end{equation}

For later discussions, it is convenient to define an average transport neutrinosphere~\cite{Raffelt:2001} $R_\text{tr}$ from the condition:
\begin{equation}
  \label{eq:tautrans}
  \tau_\text{tr} (R_\text{tr}) = \int_{R_\text{tr}}^\infty dr \left\langle
    \frac{1}{\lambda_{\nu,\text{tr}}}\right\rangle \equiv \frac{2}{3} 
\end{equation}
where the total average transport mean free path is defined as follows
\begin{equation}
  \label{eq:lambdaT}
  \left\langle\frac{1}{\lambda_{\nu,\text{tr}}}\right\rangle =
  \left\langle\frac{1}{\lambda_{\nu,\text{abs}}}\right\rangle + 
      \left\langle\frac{1}{\lambda_{\nu,\text{scatt}}}\right\rangle 
\end{equation}
Note that if absorption processes dominate, $1/\lambda_{\nu,\text{tr}}\gtrsim 1/\lambda_{\nu,\text{eff}}$, while if scattering dominates $1/\lambda_{\nu,\text{tr}}>>1/\lambda_{\nu,\text{eff}}$, implying that the energy sphere is located as smaller radius (and hence higher density) than the transport sphere. As discussed by Raffelt~\cite{Raffelt:2001}, that means once neutrinos decouple form matter at the energysphere radius, $R_\text{ES}$ (eq.~\ref{eq:tautherm}), they still suffer several elastic scattering events until they reach the transport sphere at larger radius $R_\text{tr}$. The distance between both spheres defines a scattering atmosphere.

Based on the above definitions, we will illustrate in the following the situation in the proto-neutron star (PNS) atmosphere. Fig.~\ref{fig:decoupling} shows radial profiles of matter density (solid black line) and neutrino luminosities (red lines), which correspond to a typical post-bounce time between 1--2~seconds after the onset of explosion (depending on the progenitor model). The energy sphere for $\nu_e$, see expression~\eqref{eq:tautherm} is shown by a vertical blue line in the upper part of the plot. It is located near the PNS surface where the density drops over several orders of magnitude. Note that the density gradient at the PNS surface continues to steepen during the ongoing PNS deleptonization, which will be further discussed in section~III.  To illustrate the dependence of neutrino decoupling on energy we also show the energy dependent neutrinospheres, where color coding is according to the neutrino energy between 3~MeV (black) to 300~MeV (white). Low energy neutrinos of 3--5~MeV decouple already at high densities ($\sim 10^{14}$~g~cm$^{-3}$), while high-energy neutrinos (100--300~MeV) decouple at rather low densities ($\sim 10^8$~g~cm$^{-3}$). The luminosities can vary until the final neutrino decoupling radius, outside which the neutrino luminosities stay constant and hence the spectra are frozen.

Note that the luminosity hierarchy shown in Fig.~\ref{fig:decoupling} and its detailed evolution over time depends on the neutrino opacities used (see for example~\cite{Huedepohl:2010}), however, as the proto-neutron star deleptonizes one observes that both the luminosities and average energies of all neutrino flavors converge to almost indistinguishable values. It is one of the objectives of this manuscript to address the physical origin of this behavior.

\section{O-Ne-Mg-core supernova explosion}

We begin considering the 8.8~M$_\odot$ O-Ne-Mg-core collapse simulations. The model was explored consistently though core collapse, bounce, explosion and proto-neutron star (PNS) deleptonization up to 7~seconds after core bounce~\cite{Fischer:2009af}. The early onset of explosion for this model, at already 35~ms post bounce, is related to the special structure of the progenitor that lacks an extended high-density envelope. For details about the explosion mechanism, see refs.~\cite{Kitaura:2006,Fischer:2009af}. We will start our analysis by summarizing the evolution of neutrino luminosities and mean energies. Further below we will examine important aspects related to the proto-neutrons star (PNS) deleptonization.

\subsection{Proto-neutron star deleptonization}

\begin{figure}
\includegraphics[width=0.45\textwidth]{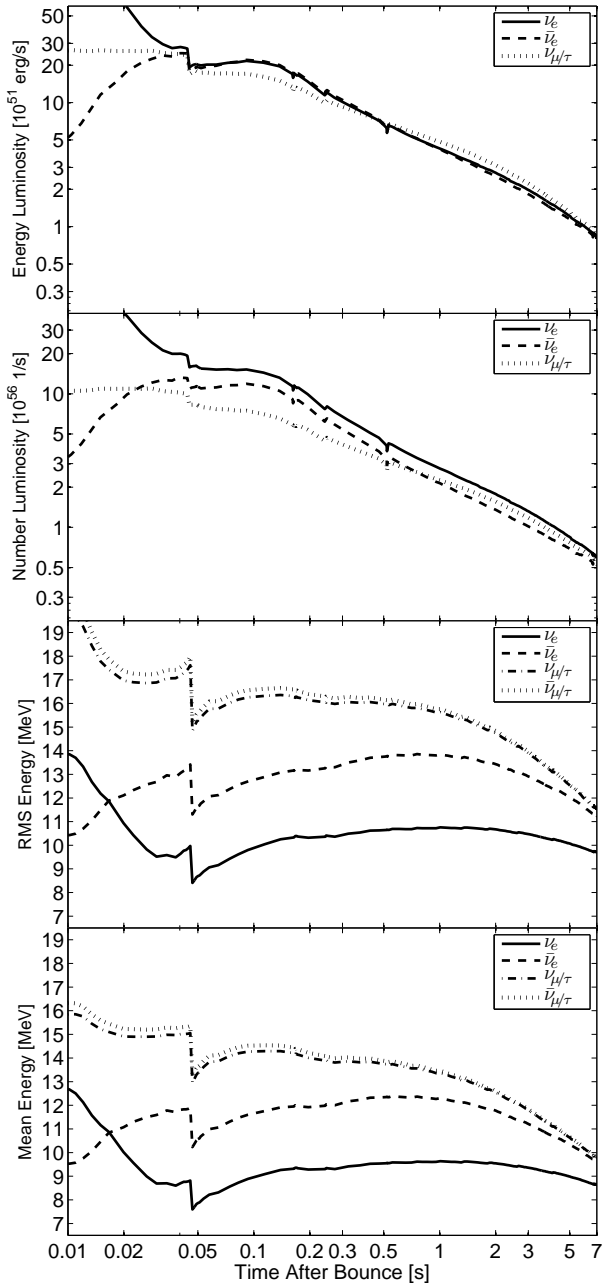}
\caption{Post-bounce evolution of neutrino energy and number luminosities as well as mean and root-mean-square (rms) energies for the 8.8~M$_\odot$ O-Ne-Mg-core supernova~\cite{Fischer:2009af}. We show explicitly the average energies for neutrinos and antineutrinos of all flavors. For the luminosities, we only show $\nu_e$, $\bar\nu_e$ and $\nu_{\mu/\tau}$ because $\bar\nu_{\mu/\tau}$ cannot be distinguished from $\nu_{\mu/\tau}$ at the scale chosen.}
\label{fig:lumin-n08c}
\end{figure}

Fig.~\ref{fig:lumin-n08c} shows neutrino energy $L_\nu$ and number luminosities $L_{n,\nu}$ as well as root-mean-square and mean energies, with respect to time after bounce (in log-scale). For a definition of these quantities, see ref.~\cite{Liebendoerfer:2005a}. All plotted quantities are measured in the co-moving reference frame at a distance of 500~km from the center (see ref.~\cite{Fischer:2009af}). The sharp luminosity and energy jumps at about 40~ms post bounce are related to Doppler shift effects, where due to the passage of the supernova shock, matter velocities suddenly change from infall to expansion. During the early post-bounce phase until 40~ms, the electron neutrino luminosities are dominated by mass accretion at the neutrinospheres. They reach values of about $2\times10^{52}$~erg/s and stay relatively constant during the accretion phase, after the deleptonization burst from the core bounce has been launched. Note that during the accretion phase, the electron neutrino energy and number luminosities are larger than the electron antineutrino energy and number luminosities. However, during the early explosion phase the electron antineutrino energy luminosity becomes slightly larger than the electron neutrino energy luminosity (see Fig.~\ref{fig:lumin-n08c}). The following hierarchy holds $L_{\bar\nu_e} \gtrsim L_{\nu_e} > L_{\nu_{\mu/\tau}}$ until about $350$~ms post bounce, after which the electron flavor neutrino luminosities change ordering and the following hierarchy holds $L_{\nu_{\mu/\tau}} > L_{\bar\nu_e} \gtrsim L_{\bar\nu_e}$. This hierarchy remains until the simulation is stopped at about 7~seconds post bounce. The luminosities of all flavors decrease by about one order of magnitude within the first second after the onset of explosion. For the number luminosities the following hierarchy holds $L_{n,\nu_e} > L_{n,\bar{\nu}_e} > L_{n,\nu_{\mu/\tau}}$ up to about $750$~ms post bounce, after which $L_{n,\nu_e} > L_{n,\nu_{\mu/\tau}} > L_{n,\bar{\nu}_e}$. This continuous decrease of the neutrino luminosities reflects the ongoing proto-neutron star deleptonization.

Of additional relevance for nucleosynthesis and flavor oscillation studies are the mean energies, shown in Fig.~\ref{fig:lumin-n08c} (bottom), in particular the difference between $\nu_e$ and $\bar\nu_e$. During the accretion phase before the onset of an explosion, $\langle E_{\nu_{\mu/\tau}}\rangle$ decrease slightly from 16~MeV to 15~MeV, $\langle E_{\bar\nu_e}\rangle$ increases from 10~MeV to 12~MeV and $\langle E_{\nu_e}\rangle$ decrease from 13~MeV to 8~MeV. After the onset of explosion, both $\langle E_{\nu_e}\rangle$ and $\langle E_{\bar\nu_e}\rangle$ increase slightly from 7.5~MeV and 10~MeV to 8.5~MeV and 12~MeV at about 350~ms post bounce. Both mean energies increase and their difference stays as large and even increases slightly. However, the difference is not large enough to turn matter neutron rich. $\langle E_{\bar\nu_{\mu/\tau}}\rangle$ decrease continuously from about 14~MeV at the onset of explosion to 8~MeV at 7~seconds post bounce. After about 1~s post bounce, also $\langle E_{\bar\nu_e}\rangle$ and $\langle E_{\nu_e}\rangle$ decrease continuously to 9~MeV and 8~MeV. The rms-energies are slightly larger than the mean energies but follow the same behavior. The decreasing mean energies for all flavors indicates the ongoing deleptonization of the central PNS and hence cooling by neutrinos.
Furthermore, the mean energies of all flavors become increasingly similar with respect to time during the PNS deleptonization (see Fig.~\ref{fig:lumin-n08c}).

\begin{figure*}
\centering
\includegraphics[width=0.95\textwidth]{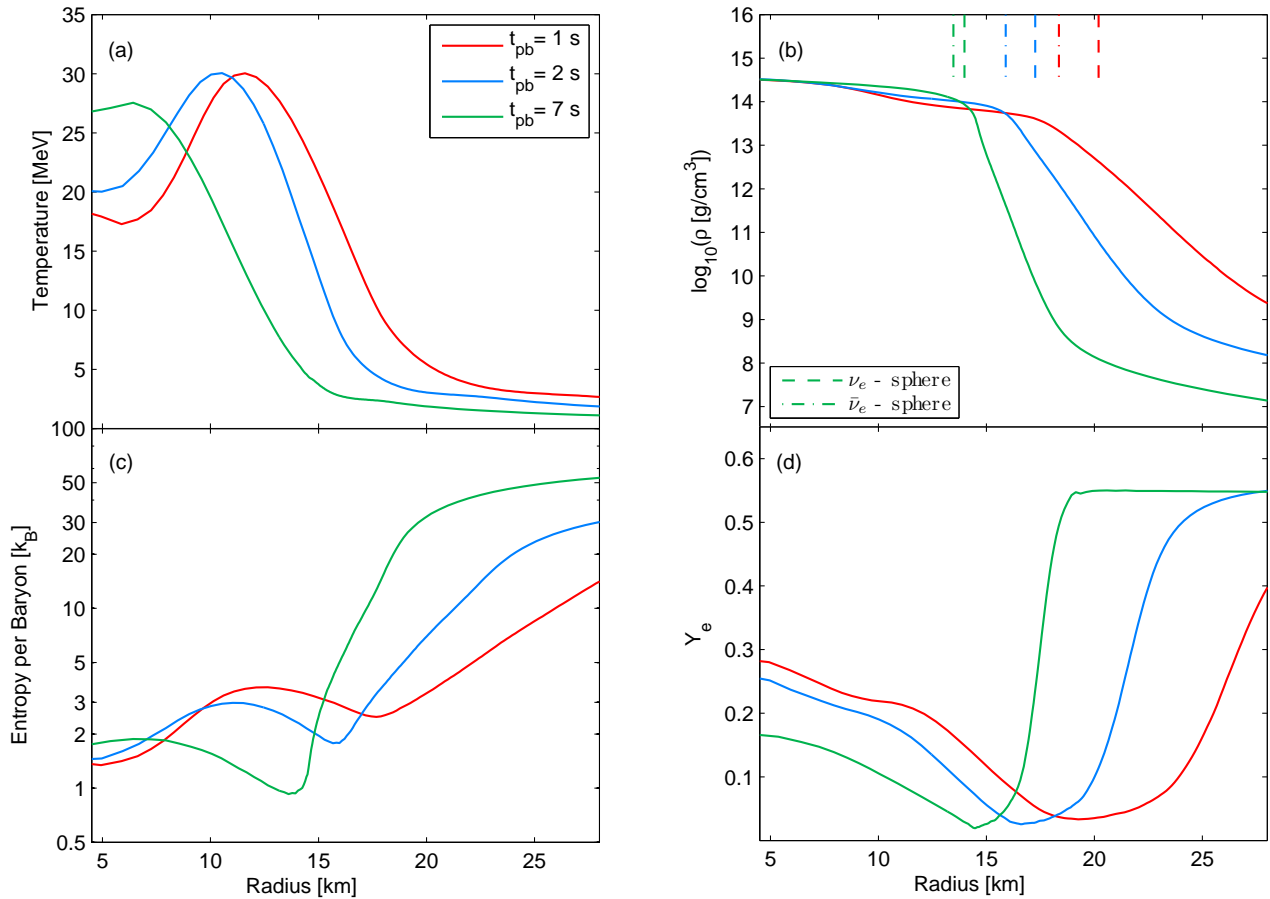}
\caption{Radial profiles for selected quantities, for the 8.8~M$_\odot$ O-Ne-Mg-core \cite{Fischer:2009af}, at three different post-bounce times (1~second (red lines), 2~seconds (blue lines), 7~seconds (green lines)). The vertical lines in the graph (b) correspond to the position of the energy spheres for  $\nu_e$ (dashed lines) and $\bar{\nu}_e$ (dash-dotted lines).} \label{fig:fullstatemoments-n08c}
\end{figure*}

The resulting evolution of the explosion and the neutrino spectra is in good qualitative and quantitative agreement with the study of the Garching group of this low-mass progenitor model~\cite{Kitaura:2006,Huedepohl:2010}, applying a different equation of state and in addition a set of updated weak interactions processes.

The evolution of radial profiles of selected quantities is illustrated in Fig~\ref{fig:fullstatemoments-n08c} at several post-bounce times (1~seconds: red lines, 2~seconds: blue lines, 7~seconds: green lines). We focus on the radial domain near the neutrinospheres (vertical dashed lines for $\nu_e$ and dash-dotted lines for $\bar\nu_e$ in graph~(b)), i.e. the region where neutrinos decouple from matter and where the far distance spectra are determined. The graphs (a), (c) and (d) show radial profiles of temperature, entropy per baryon and electron fraction, all of which decrease at the neutrinospheres. This evolution is typical for the PNS deleptonization and neutrino cooling including the slow proto-neutron star contraction. Note the rapidly rising electron fraction outside the neutrinospheres, which is related to the expansion of material in the neutrino-driven wind where $Y_e\simeq0.56$ (see ref.~\cite{Fischer:2009af} for a discussion). As the temperature reduces, the mean energy of neutrinos also decreases and the neutrinospheres move to higher densities and hence smaller radii, during the proto-neutron star deleptonization (see Fig.~\ref{fig:fullstatemoments-n08c} graphs~(a) and (b)), from $R_{\nu_e}=22.19$~km and $R_{\bar\nu_e}=21.51$~km at 1~second post bounce to $R_{\nu_e}=15.28$~km and $R_{\bar\nu_e}=14.97$~km at 7~seconds post bounce.

In the following subsection, we will analyze the reason for the decreasing difference in the mean neutrino energies.

\begin{figure*}
\subfigure[$\,\,\,\,1$~second after bounce]
{\includegraphics[width=0.85\textwidth]{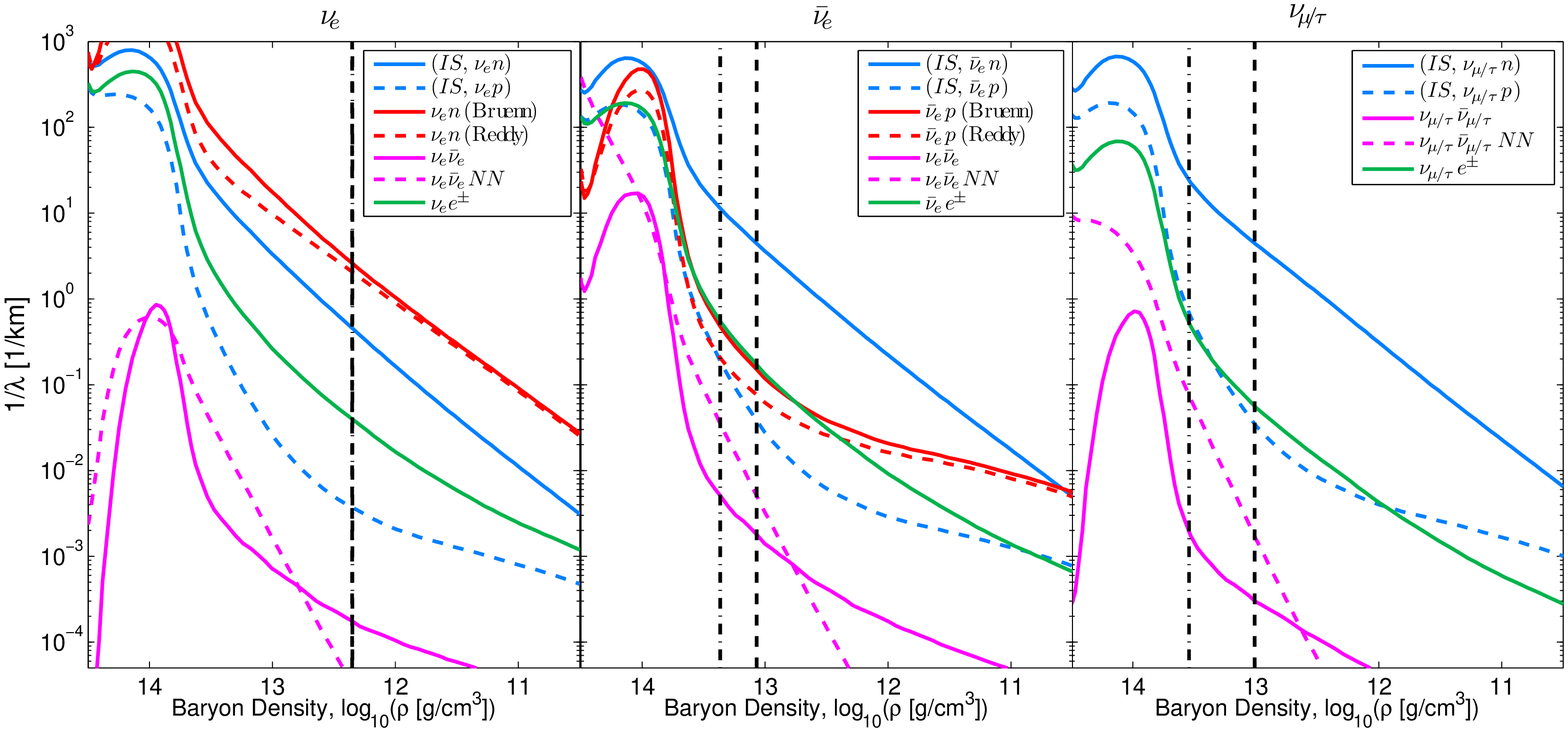}
\label{fig:mfp-n08c-a}}\\
\subfigure[$\,\,\,\,2$~seconds post bounce]
{\includegraphics[width=0.85\textwidth]{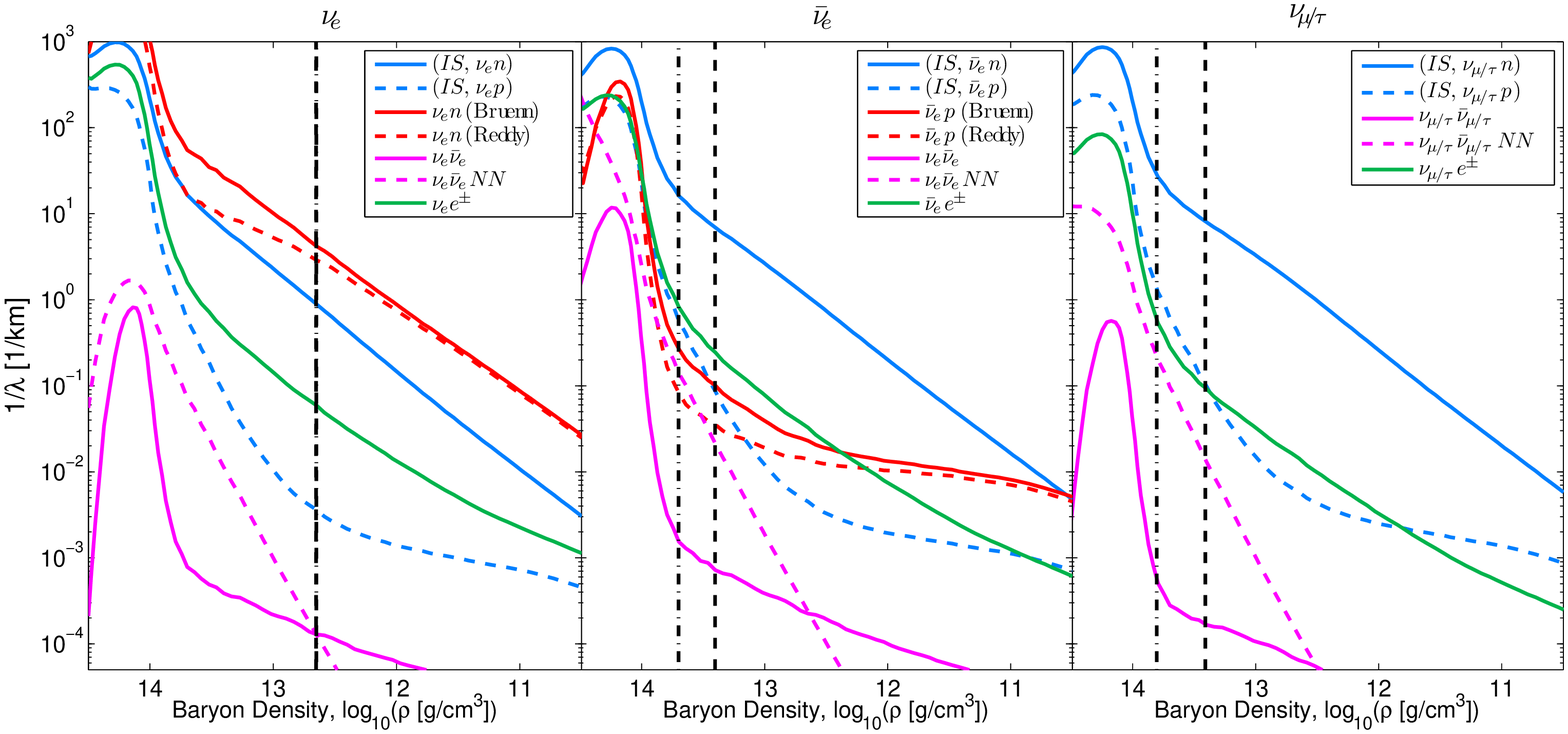}
\label{fig:mfp-n08c-b}}\\
\subfigure[$\,\,\,\,7$~seconds after bounce]
{\includegraphics[width=0.85\textwidth]{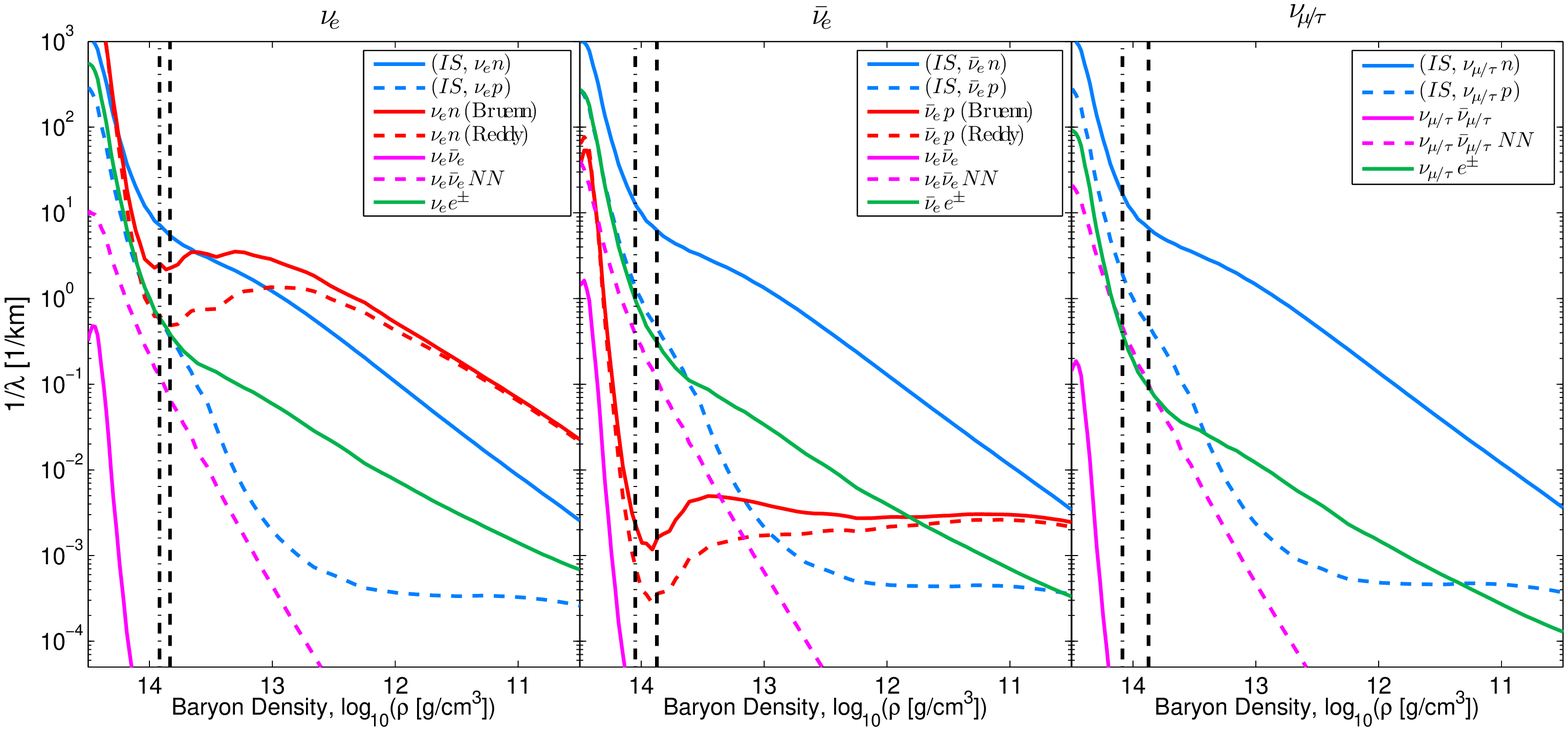}
\label{fig:mfp-n08c-c}}
\caption{Inverse mean free paths for the individual reactions considered $(\text{IS}, \nu N)$: isoenergetic neutrino-nucleon scattering, ($\nu_en$, $\bar\nu_ep$): charge-current reactions (comparing rates form ref.~\cite{Bruenn:1985en} (Bruenn) and ref.~\cite{Reddy:1998} (Reddy)), ($\nu\bar\nu$): $e^-$ $e^+$ - annihilation, ($\nu\bar\nu N N$): $N$--$N$--Bremsstrahlung), for $\nu_e$ (left panel), $\bar\nu_e$ (right panel) and $\nu_{\mu/\tau}$ (right panel), based on radial profiles and evolution of the 8.8~M$_\odot$ O-Ne-Mg-core collapse supernova simulation~\cite{Fischer:2009af}. We show selected post-bounce times 1~s (top), 2~s (middle) and 7~s (bottom). Note the different density scale for the bottom panel. The vertical black dashed and dash-dotted lines mark the position of the transport (\ref{eq:tautherm}) and energy spheres (\ref{eq:tautrans}).}
\label{fig:mfp-n08c}
\end{figure*}

\subsection{Individual opacities}

Fig.~\ref{fig:mfp-n08c} shows radial profiles of inverse mean free paths for the individual reactions considered, for $\nu_e$ (left panel), $\bar\nu_e$ (middle panel) and $\nu_{\mu/\tau}$ (right panel) at selected post-bounce times obtained during the PNS deleptonization.

We will start analyzing the inverse mean free paths for $(\mu,\tau)$-neutrinos (right panel in Fig.~\ref{fig:mfp-n08c}). Note that they have no contributions from charge-current processes, they are only produced via the neutral-current pair-creation reactions (7) and (8) in Table~\ref{table-nu-reactions}. The dominating inelastic contribution  comes from $N$--$N$--Bremsstrahlung, only a tiny contribution comes from $e^-$--$e^+$-annihilation, and scattering on  electrons/positrons.  All inelastic processes are smaller by several orders of magnitude than elastic scattering on  neutrons (IS, $\nu_{\mu/\tau}n$). Note further that elastic scattering on protons (IS, $\nu_{\mu/\tau}p$) is also smaller than scattering on neutrons because protons are much less abundant than neutrons.

The large scattering dominance implies that the transport opacity is greater than the effective opacity, $1/\lambda_{\nu_{\mu/\tau},\text{tr}} \gg 1/\lambda_{\nu_{\mu/\tau},\text{eff}}$, see equations~(\ref{eq:lambdamuE}) and (\ref{eq:lambdaT}).The corresponding neutrinospheres spheres are shown in Fig.~\ref{fig:mfp-n08c} (energy sphere: vertical black dash-dotted line, transport sphere: vertical black dashed line). The separation of both defines the scattering atmosphere for $\nu_{\mu/\tau}$, which is present already at early times after the onset of an explosion.

The situation remains the same at later times for $(\mu,\tau)$-neutrinos. However, the difference between $1/\lambda_{\text{IS}, \nu_{\mu/\tau}p}$ and $1/\lambda_{\nu_{\mu/\tau}e^\pm}$ increases with $1/\lambda_{\nu_{\mu/\tau}\bar\nu_{\mu/\tau}NN}$ becoming as large as $1/\lambda_{\nu_{\mu/\tau}e^\pm}$, while $1/\lambda_{\nu_{\mu/\tau}\bar\nu_{\mu/\tau}}$ continuously decreases with times (see the right panels Figs.~\ref{fig:mfp-n08c-a}--\ref{fig:mfp-n08c-c}).

The situation for $\bar\nu_e$ is illustrated in the middle panels of Fig.~\ref{fig:mfp-n08c}. Here, in addition to the neutral-current inelastic scattering and pair-creation processes, absorption on protons contributes to the energy exchange processes. However, their contribution is small and comparable to neutrino scattering on electrons/positrons at early times. The small value of $1/\lambda_{\bar\nu_e p}$, is related to the low number density of protons in the neutron-rich environment. At late times $1/\lambda_{\bar\nu_e p}$ even decreases an order of magnitude below $1/\lambda_{\nu_{\mu/\tau}\bar\nu_{\mu/\tau} N N}$. As the density at the energy sphere, $R_{ES}$, increases with time, the opacity for $\bar\nu_e$ absorption on protons decreases due to Pauli-blocking of final-state neutrons. This aspect is taken into account via nucleon-degeneracy factors in the charge-current reactions rates~\cite{Bruenn:1985en}.

As before for $\nu_{\mu/\tau}$, the dominating process is here also $\bar\nu_e$ scattering on neutrons. Hence similar to $(\mu/\tau)$-neutrinos, $1/\lambda_{\bar\nu_e,\text{tr}}>1/\lambda_{\bar\nu_e,\text{ eff}}$ already early after the onset of an explosion and a scattering atmosphere has developed. At early times, due to contributions from $\bar\nu_e$ absorption on protons the energy sphere for $\bar\nu_e$ is located at larger radius, i.e. at lower matter temperature, than the energy sphere for $\nu_{\mu/\tau}$. However, as the opacity for $\bar\nu_e$ absorption on protons decreases with time the temperatures at the energy spheres of $\bar\nu_e$ and $\nu_{\mu/\tau}$ become increasingly similar. This explains the convergence of mean energies and luminosities for $\bar\nu_e$ and $\nu_{\mu/\tau}$ at late times (see Fig.~\ref{fig:lumin-n08c}).

Note the constant opacity for $\bar\nu_e$ absorption on protons outside the transport sphere at 7~seconds post bounce in Fig.~\ref{fig:mfp-n08c}, which is due to the constant number density of protons where the increasing abundance of protons is compensated by the decreasing density at the proto-neutron star surface.

In contrast to $\bar\nu_e$ and $\nu_{\mu/\tau}$, the situation is different for $\nu_e$ illustrated at the left panels in Fig.~\ref{fig:mfp-n08c}. The opacity for $\nu_e$ is initially dominated by charge current $\nu_e$ absorption on neutrons, $1/\lambda_{\nu_e n}$. It is greater than $1/\lambda_{\text{IS}, \nu_e n}$ by several orders of magnitude early after the onset of explosion at 1~second post bounce, see Fig.~\ref{fig:mfp-n08c-a}. As the total opacity is dominated by absorption processes, we are in a situation where $1/\lambda_{\nu_e,\text{tr}} \gtrsim 1/\lambda_{\nu_e,\text{ eff}}$. Hence, a scattering atmosphere is not present for $\nu_e$ a few seconds after the onset of an explosion~\cite{Raffelt:2001}. However, around 7~seconds post bounce a scattering atmosphere starts to develop, see Fig.~\ref{fig:mfp-n08c-c}. Similar as for $\bar\nu_e$, the opacity for neutrino absorption on neutrons becomes smaller than the one for neutrino scattering on neutrons in the region around the neutrinospheres. It is due to Pauli blocking of final-state electrons.

At the time when the simulations were stopped, the effective opacity of $\nu_e$ still contains substantial contributions from charge-current processes. Hence the spectra of $\nu_e$ and $\bar\nu_e$, the latter contains negligible contributions from charge-current contributions, have not converged yet. As will be discussed in the next section, we expect full convergence of the neutrino spectra and luminosities for $\nu_e$ and $\bar\nu_e$ at later times.

Note that for previous analysis is based on charge-current rates from \cite{Bruenn:1985en} computed in the so called elastic approximation, which assumes zero-momentum transfer. This assumption may not be accurate at high densities, where neutrons become degenerate and neutrinos have higher average energies. To check this issue, we have included the full energy and momentum-transfer dependent treatment of ref.~\cite{Reddy:1998} for the charge-current reactions. The corresponding opacity based on that treatment is shown in Fig.~\ref{fig:mfp-n08c} (red dashed lines, labelled 'Reddy') for $\nu_e$ and $\bar\nu_e$, in comparison to the opacity form ref.~\cite{Bruenn:1985en} (red solid lines, labelled 'Bruenn'). At low densities, the differences between both opacities are minor. At high densities the differences between both approaches increases at late times as matter becomes increasingly neutron rich and the temperature decreases, which in turn rises the neutron degeneracy. 

It indicates that Pauli-blocking of final states described above, relevant for the suppression of charge-current reactions during the deleptonization, will even increase in the simulations when applying the improved treatment of ref.~\cite{Reddy:1998}.

\begin{figure}[htp!]
\includegraphics[width=\columnwidth]{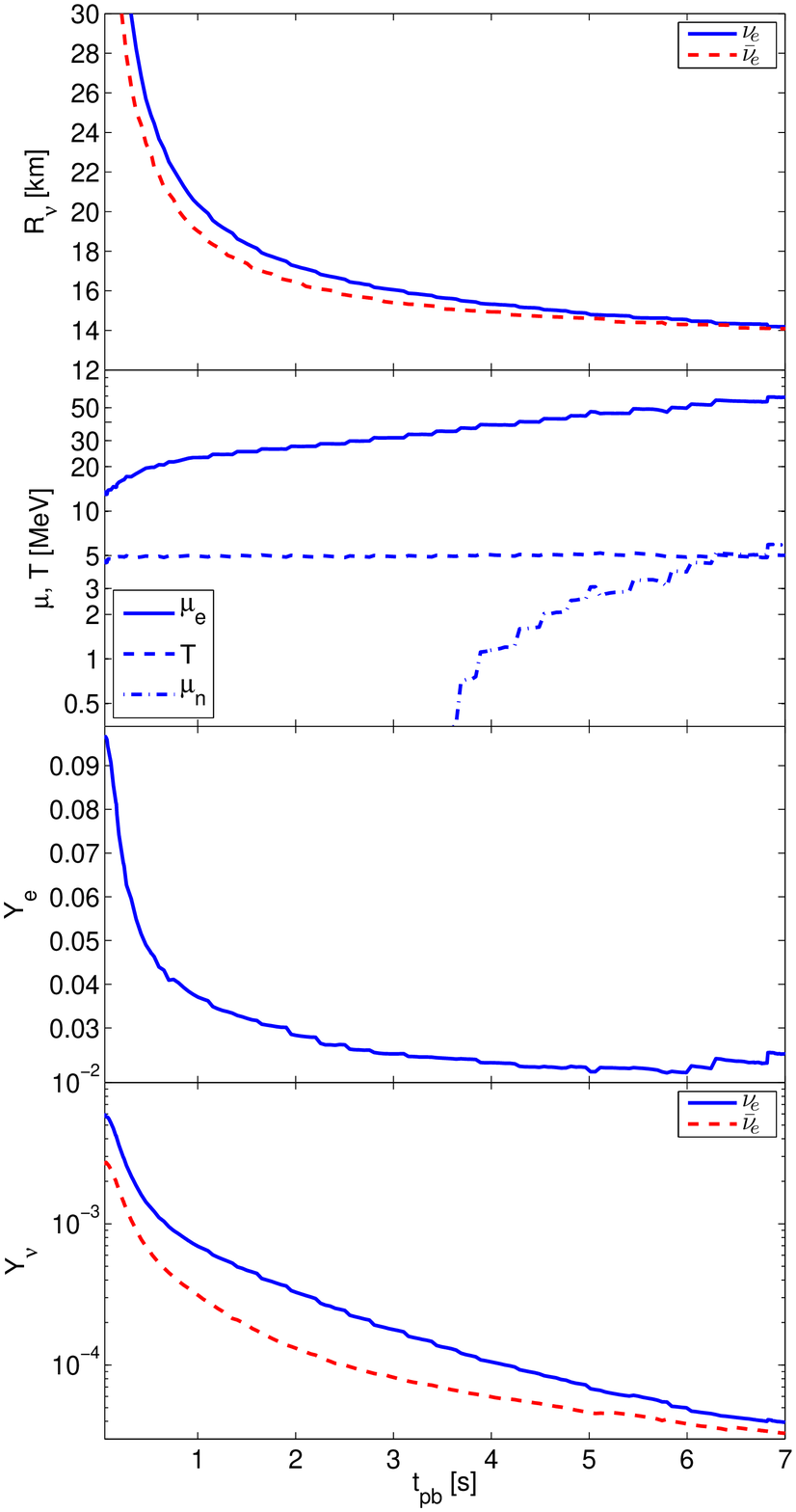}
\caption{Post-bounce evolution of the energy spheres for $\nu_e$ and $\bar\nu_e$, electron and neutron chemical potentials as well as temperature and electron fraction at the $\nu_e$-sphere. Furthermore, the evolution of the $\nu_e$ and $\bar\nu_e$ fractions is shown at the corresponding spheres.}
\label{fig:delept-n08c}
\end{figure}
%

Now, we discuss the evolution of selected quantities at the $\nu_e$ energy sphere illustrated in Fig.~\ref{fig:delept-n08c} during the deleptonization. Panel~(a) shows the evolution of the sphere radii for $\nu_e$ (solid line) and $\bar\nu_e$ (dashed line). The temperature evolution at the $\nu_e$-sphere is shown in Fig.~\ref{fig:delept-n08c}~(b), which remains about constant at  5~MeV. Fig.~\ref{fig:delept-n08c}~(c) and (d) show the evolution of electron fraction at the $\nu_e$-sphere and also the $\nu_e$ and $\bar\nu_e$ fractions at the corresponding spheres. The electron fraction drops rapidly from 0.1 to 0.035 during the first second after the onset of  explosion. At later times, the electron fraction evolves to saturation values of 0.03 (at $R_{ES}$ for $\nu_e$) and 0.025 (at $R_{ES}$ for $\bar\nu_e$). The neutrino fractions are already very small inside the proto-neutron star at the onset of explosion, with continuously decreasing values $Y_{\nu_e}\simeq 10^{-2}$ and $Y_{\bar\nu_e}\simeq 4\times10^{-3}$ at the corresponding spheres. Later during the deleptonization, the neutrino fractions continuously decrease to $Y_{\nu_e}\simeq 3\times10^{-5}$ and $Y_{\bar\nu_e}\simeq 2\times10^{-3}$ at about 7~seconds post bounce.
Furthermore, we also show the evolution of the neutron chemical potential, $\mu_n$, at the $\nu_e$-sphere in Fig.~\ref{fig:delept-n08c}~(b). It rises continuously and reaches positive values at about 3.5~seconds post bounce, after which the neutrons have Fermi energies greater than their restmass.
$\mu_n$ exceeds the temperature at about 6~seconds post bounce, after which the neutrons become degenerate. Equally important is the evolution of the electron chemical potential, $\mu_e$, at the $\nu_e$-sphere, which is shown in Fig.~\ref{fig:delept-n08c}~(b) too. It rises from around 10~MeV to 60~MeV during deleptonization, while the temperature decreases only slightly. The rising electron chemical potential together with the decrease of the mean neutrino energy implies increasing importance of final-state electron blocking.

\begin{figure}
\includegraphics[width=0.45\textwidth]{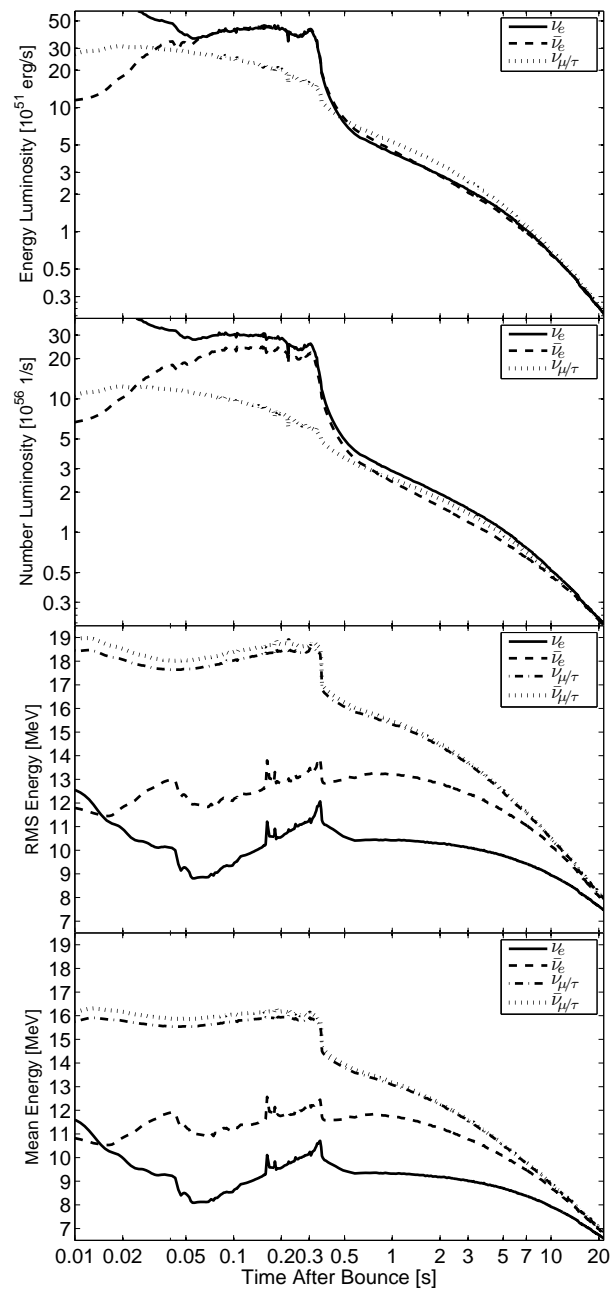}
\caption{Post-bounce evolution of neutrino energy and number luminosities as well as mean and rms-energies for the 18~M$_\odot$ Fe-core progenitors from ref.~\cite{Fischer:2009af}.}
\label{fig:lumin-h18b}
\end{figure}
%

\section{Iron-core progenitors}

\begin{figure*}
\centering
\includegraphics[width=0.95\textwidth]{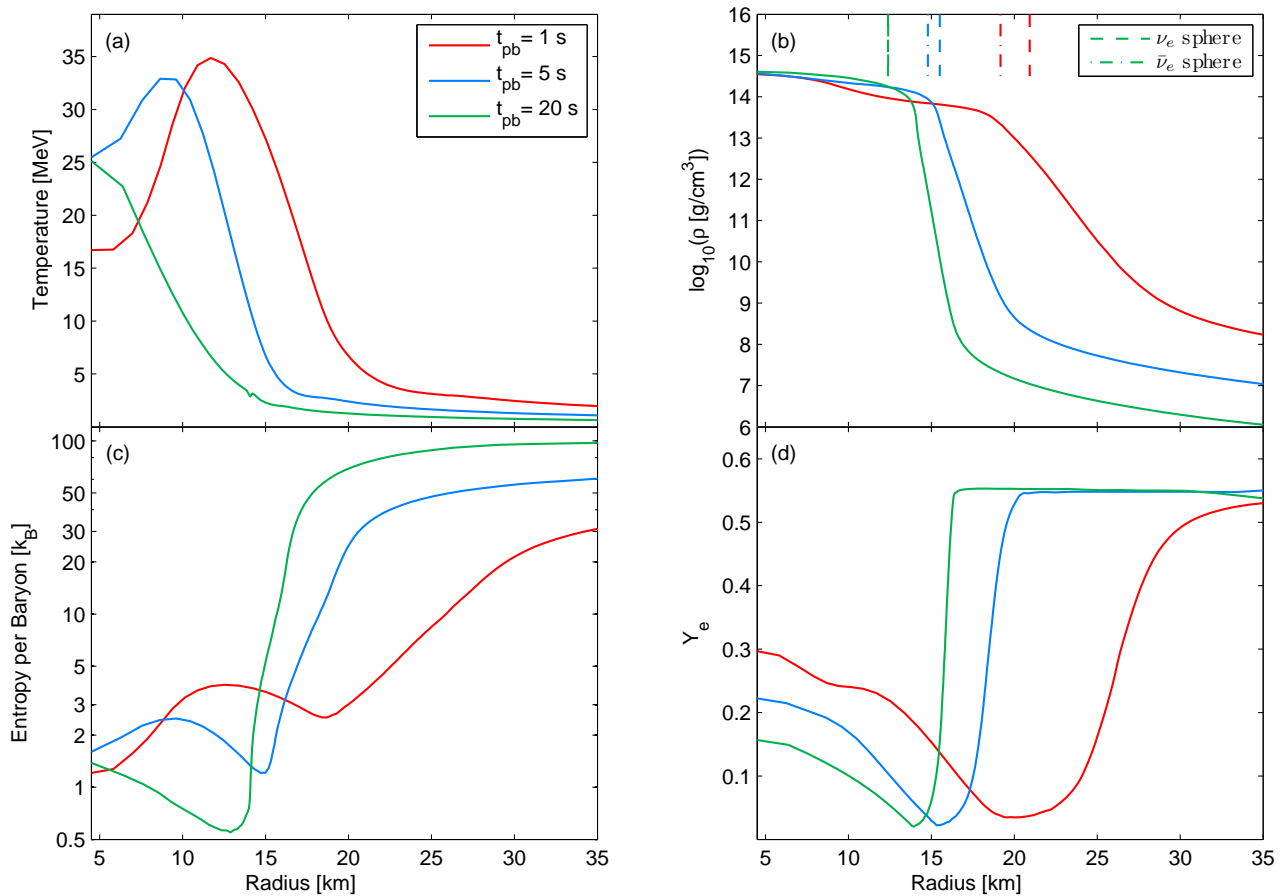}
\caption{Radial profiles of selected hydrodynamic variables at $1$~seconds (red lines), at $5$~seconds (blue lines) and 20~seconds (green lines) post bounce. The same configuration as Fig.~\ref{fig:fullstatemoments-n08c} but for the 18~M$_\odot$ progenitor model from ref.~\cite{Fischer:2009af}.}
\label{fig:fullstatemoments-h18b}
\end{figure*}

In this section, we will compare the results found above for the low-mass O-Ne-Mg-core collapse supernova simulation with the more massive 18~M$_\odot$ iron-core progenitor~\cite{Woosley:2002zz}. Note that neutrino-driven explosions of stellar models other than the O-Ne-Mg-core cannot be obtained using the standard physics input in spherical symmetry. Hence, in order to trigger the explosion for the 18~M$_\odot$ progenitor under investigation, neutrino absorption and emission rates were enhanced in the gain region. Here we focus on the cooling phase of the proto-neutron star after the onset of explosion, that is not affected by the enhanced rates used to trigger the explosion. We have checked this by doing reference simulations without enhanced rates where differences obtained are less than 5\%, compared to the results presented in this section.

Note further that in contrast to the O-Ne-Mg-core model, which was evolved up to 7~seconds post bounce where the spectra of $\nu_e$ and $\bar\nu_e$ have not fully converged yet, the 18~M$_\odot$ model was evolved for more than 20~seconds post bounce. Here the spectra have further converged and are indistinguishable for practical applications.

\begin{figure*}
\subfigure[$\,\,\,\,1$~second after bounce]
{\includegraphics[width=0.85\textwidth]{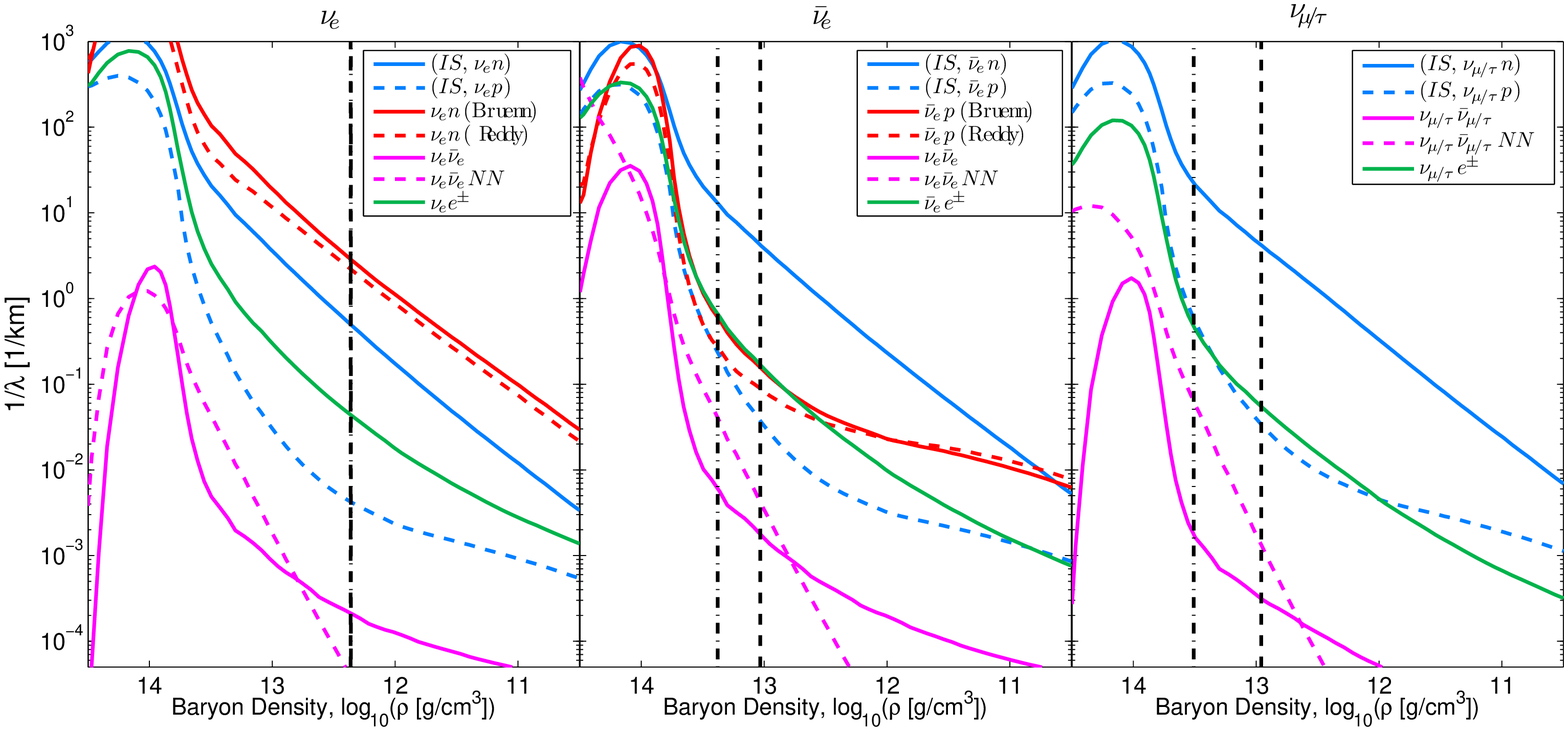}}\\
\subfigure[$\,\,\,\,5$~seconds post bounce]
{\includegraphics[width=0.85\textwidth]{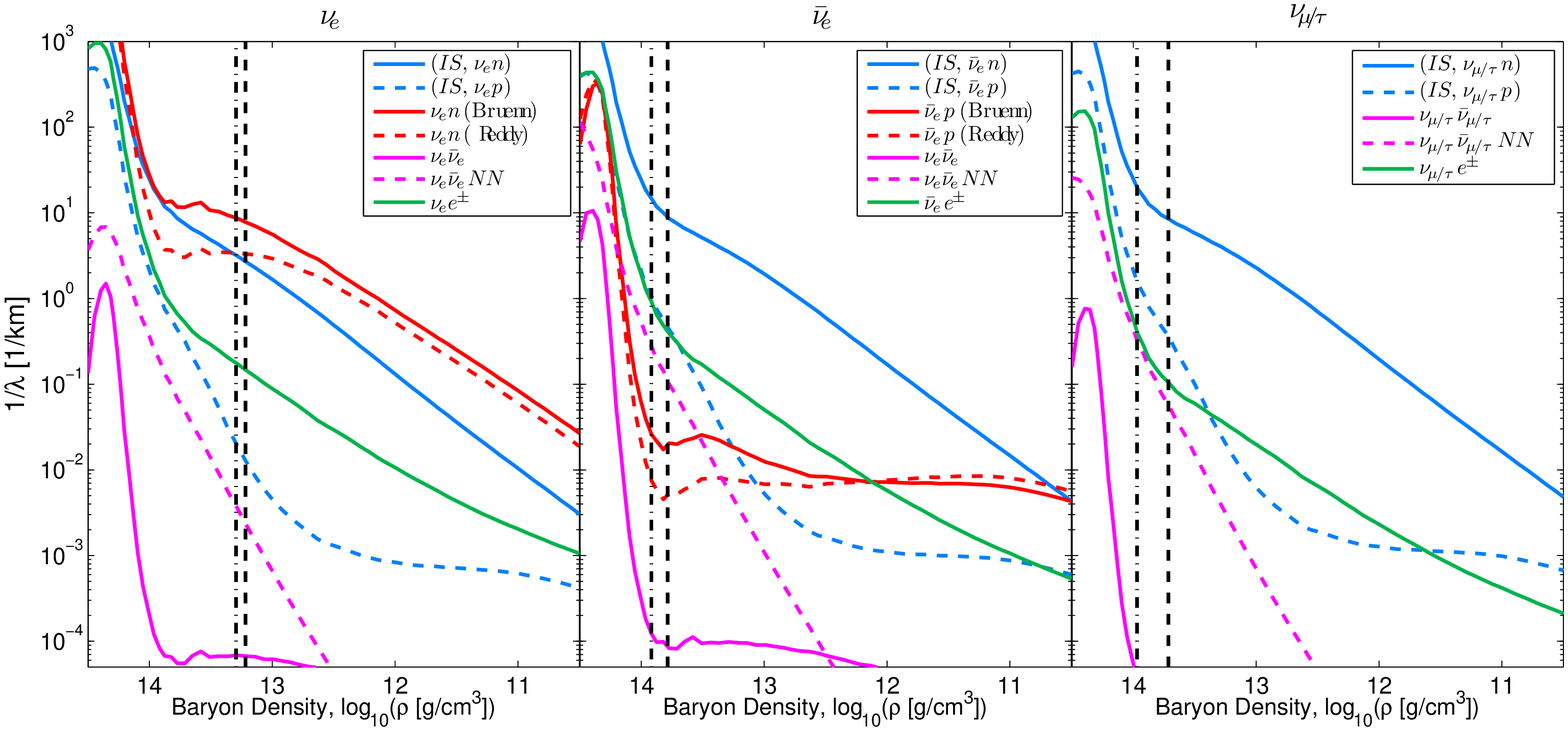}}\\
\subfigure[$\,\,\,\,20$~seconds after bounce]
{\includegraphics[width=0.85\textwidth]{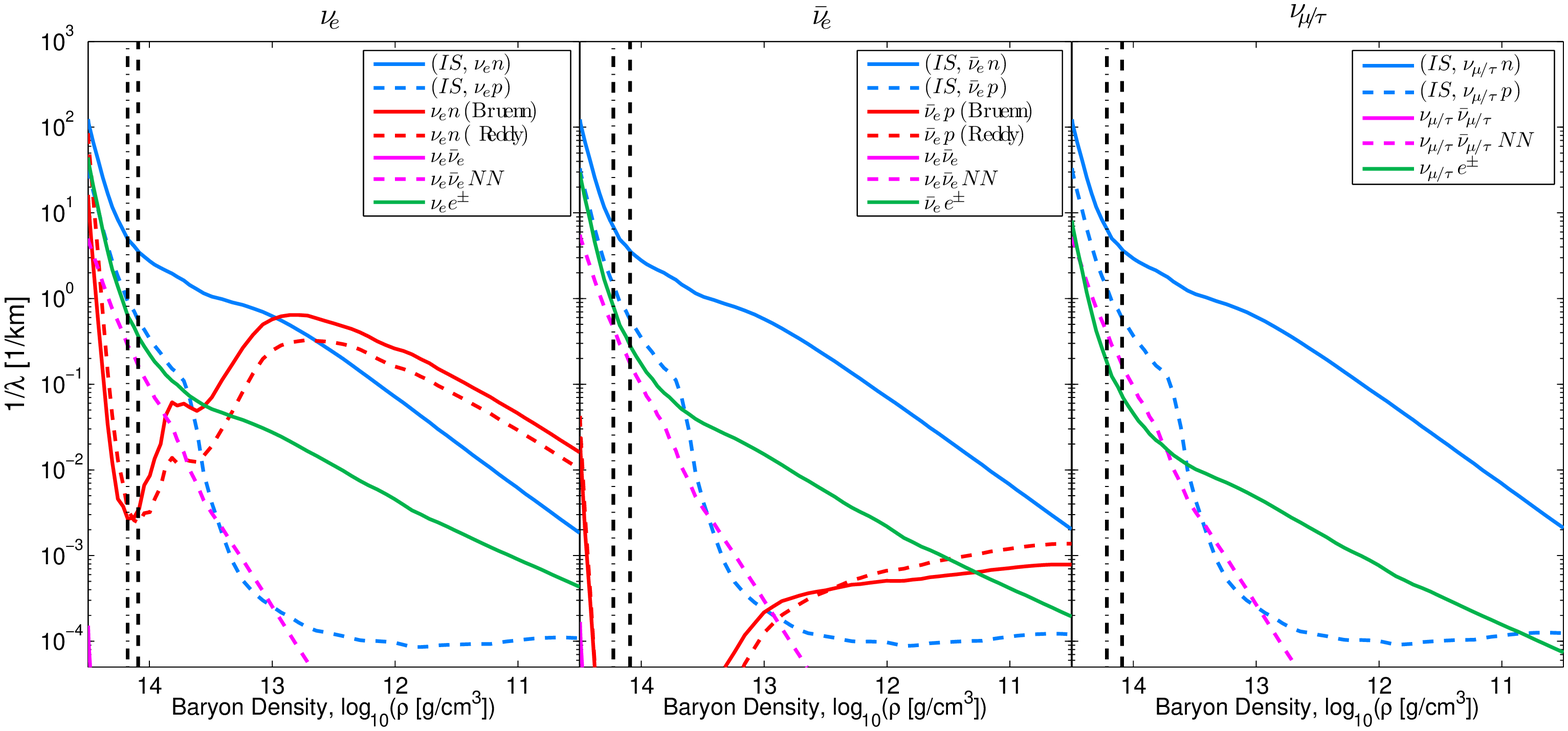}}
\caption{Individual inverse mean free paths for the weak processes considered, based on the 18~M$_\odot$ iron-core model \cite{Fischer:2009af}, at selected post-bounce times (1~second: red, 5~seconds: blue, 20~seconds: green). Vertical dashed and dash-dotted lines show the positions of transport and energy spheres. The same configuration as Fig.~\ref{fig:mfp-n08c}.}
\label{fig:mfp-h18b}
\end{figure*}

Fig.~\ref{fig:lumin-h18b} shows the evolution of the neutrino observables for the 18~M$_\odot$ model, again measured in a co-moving frame of reference at a distance of 500~km. The sharp jumps in luminosities and mean energies at about 350~ms post bounce are due to the onset of the explosion, where mass accretion vanishes and matter velocities suddenly change from infall to expansion (compare with the O-Ne-Mg-core in Fig.~\ref{fig:lumin-n08c} where the explosion is launched already at about 35~ms post bounce). The post bounce mass accretion phase lasts for several 100~ms for this iron-core progenitor. Consequently the central PNS is more compact than for the O-Ne-Mg-progenitor, with higher central density and temperature as well as lower proton-to-baryon ratio. Furthermore, the longer accretion phase and the more compact PNS lead to very similar $\nu_e$ and $\bar\nu_e$ energy luminosities during the accretion phase (see Fig.~\ref{fig:lumin-h18b}), with the same ordering $L_{\bar\nu_e} \geq L_{\nu_e} \gg L_{\nu_{\mu/\tau}}$. The number luminosities follow the same ordering here as for the O-Ne-Mg core, $L_{n,\nu_e}> L_{n, \bar\nu_e} > L_{n, \nu_{\mu/\tau}}$. The opposite ordering holds for the mean energies, $\langle E_{\nu_e}\rangle<\langle E_{\bar\nu_e} \rangle \ll \langle E_{\nu_{\mu/\tau}} \rangle \leq \langle E_{\bar\nu_{\mu/\tau}} \rangle$. They slowly rise during the accretion phase from $\langle E_{\nu_e}\rangle=8$~MeV to 10~MeV, $\langle E_{\bar\nu_e}\rangle=11$~MeV to 12.5~MeV, $\langle E_{\nu_{\mu/\tau}} \rangle=15.5$~MeV to 16~MeV at the onset of explosion.

At about 350~ms post bounce, the luminosities decrease rapidly from $4 \times 10^{52}$~erg/s ($\nu_e$, $\bar\nu_e$) and $1.5\times10^{52}$~erg/s ($\nu_{\mu/\tau}$) to $7\times 10^{51}$~erg/s (for all flavors) at about 0.5~seconds post bounce. They also change ordering $L_{\nu_{\mu/\tau}} > L_{\bar\nu_e} \simeq L_{\nu_e}$. The same holds for the number luminosities, $L_{n,\nu_e}> L_{n, \nu_{\mu/\tau}} > L_{n, \bar\nu_e}$. At about 20~seconds post bounce, the energy and number luminosities converge to $1\times 10^{50}$~erg/s and $1\times 10^{54}$~s$^{-1}$. The mean energies decrease also rapidly shortly after the onset of explosion (see bottom panels in Fig.~\ref{fig:lumin-h18b}). On timescales of several seconds, the mean energies of all flavors decrease continuously and become increasingly similar. This evolution is in qualitative agreement with evolution of the O-Ne-Mg-core collapse supernova, discussed in section III. However, for the iron-core model under discussion here, the simulation was carried out for more than 20~seconds aftre bounce. This allows us to explore with additional details the spectral convergence of the different neutrino flavors at late times.

The evolution of radial profiles of selected hydrodynamic quantities is illustrated in Fig.~\ref{fig:fullstatemoments-h18b} at the three selected post-bounce times 1, 5 and 20~seconds for the 18~M$_\odot$ iron-core progenitor model. Illustrated are conditions near the neutrinospheres during the PNS deleptonization. Similar to the 8.8~M$_\odot$ model discussed in section~III (see  Fig.~\ref{fig:fullstatemoments-n08c}) temperature, entropy per baryon and electron fraction shown in graphs~(a), (c) and (d) decrease continuously with time, while density at the neutrinospheres in graph~(b) increases. The $\nu_e$ and $\bar\nu_e$ spheres move not only to higher density but also closer together during the proto-neutron star deleptonization on timescales on the order of several seconds (see graph~(b)). It indicates the same behavior as for the O-Ne-Mg-core.

In the following we repeat the analysis from section~III B and look at the individual opacities (inverse mean free paths) at selected times during the proto-neutron star deleptonization, illustrated in Fig~\ref{fig:mfp-h18b}. Identical as for the O-Ne-Mg core, the opacity for $\nu_{\mu/\tau}$ is dominated by scattering on neutrons at any time. Inelastic processes, which are more than one order of magnitude smaller than elastic scattering on neutrons, are dominated by scattering on electrons/positrons. Only at late times after about 5~seconds post bounce, the opacity of Bremsstrahlung rises and becomes of equal importance. The positions of energy sphere (dominated by inelastic processes) and transport sphere (all processes contribute equally) are marked by the dash-dotted and dashed lines in Fig.~\ref{fig:mfp-h18b}. Their separation indicates the presence of a scattering atmosphere, which is already present for $\nu_{\mu/\tau}$ and $\bar\nu_e$ early after the onset of explosion. Similar to the O-Ne-Mg-core, the situation is different for $\nu_e$, for which the dominating opacity is absorption on neutrons. Consequently, energy and transport spheres lay close to each other and a scattering atmosphere has not yet developed. However, during deleptonization, the opacity for $\nu_e$ absorption on neutrons reduces continuously due to final state electron blocking. At about 10~seconds after bounce, the opacity from $\nu_e$ scattering on neutrons becomes larger that the opacity for absorption on neutrons, and a scattering atmosphere starts to develop. At about 20~seconds post bounce, the opacity of all flavors is dominated by neutral-current processes that do not distinguish between different flavors. The energy and transport spheres converge for all flavors and their spectra become increasingly similar. The evolution of opacities (inverse mean free paths) is in qualitative and quantitative agreement with the evolution for the O-Ne-Mg core discussed in section~III B. We also find that the improved treatment of \cite{Reddy:1998} leads to an additional reduction of the charge-current opacities (see Fig.~\ref{fig:mfp-h18b}).

Note that the extension in density of the scattering atmosphere, i.e. the density domain between energy and transport spheres, reduces during the proto-neutron star deleptonization. However, measured in radius, the scattering atmosphere increases in size.
\begin{figure*}[htp!]
\subfigure[\ 600~ms post bounce]{
\includegraphics[width=0.48\textwidth]{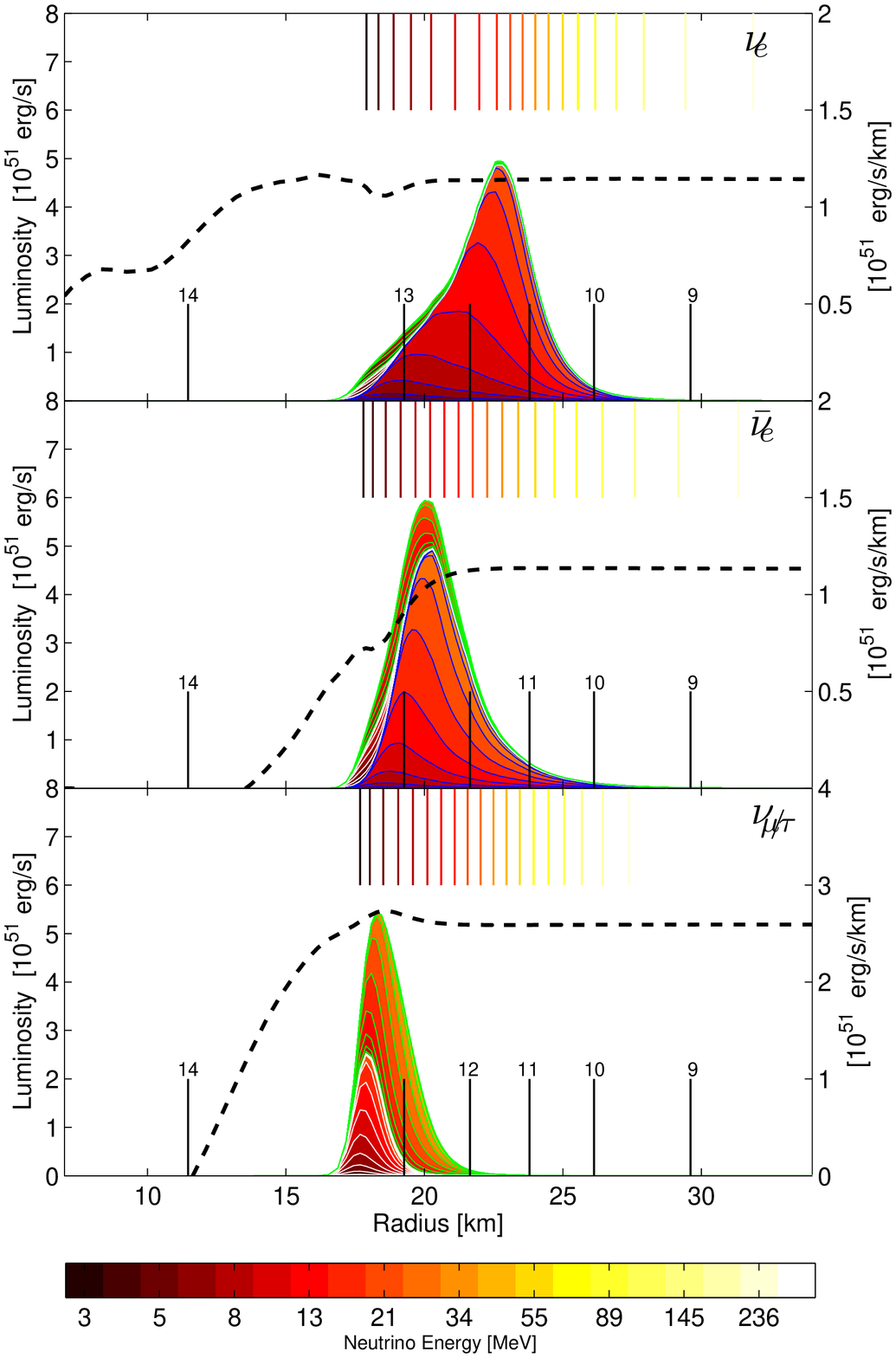}
\label{fig:nustatistics-n08c-a}}
\hfill
\subfigure[\ 7~seconds post bounce]{
\includegraphics[width=0.48\textwidth]{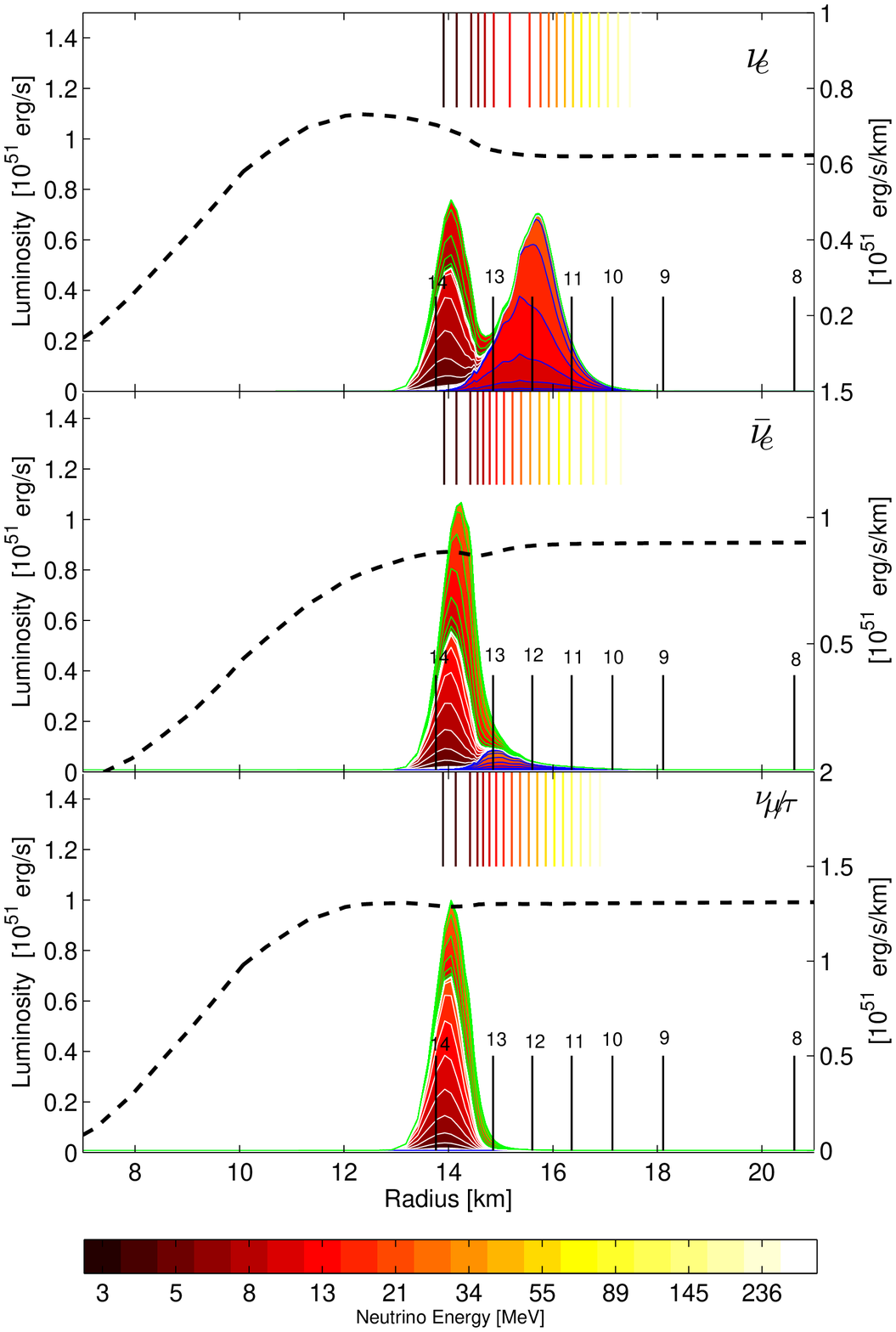}
\label{fig:nustatistics-n08c-b}}
\caption{Local contributions, $\mathcal{L}_i(E,r)$, to the luminosity at infinity (see text for its definition) at two different post-bounce times during the protoneutron star deleptonization 8.8~M$_\odot$. Note the different $y$-scales for each neutrino flavor. The color of the contour lines mark different weak processes (blue: charged current, white: pair processes, green: $\nu e^\pm$-scattering). The color gradient reflects the neutrino energy from 3 to 300 MeV (see color bar at the bottom). The dashed black lines show the luminosities and the vertical lines the position of the energy spheres (equation 13). The black vertical lines at the bottom of each panel mark different baryon densities, indicated by the labels from $\log_{10}\left(\rho\,\,\,[\text{g/cm}^3]\right)=$ 9--14.}
\label{fig:nustatistics-n08c}
\end{figure*}
\begin{figure*}[htp!]
\subfigure[\ 5~seconds post bounce]{
\includegraphics[width=0.48\textwidth]{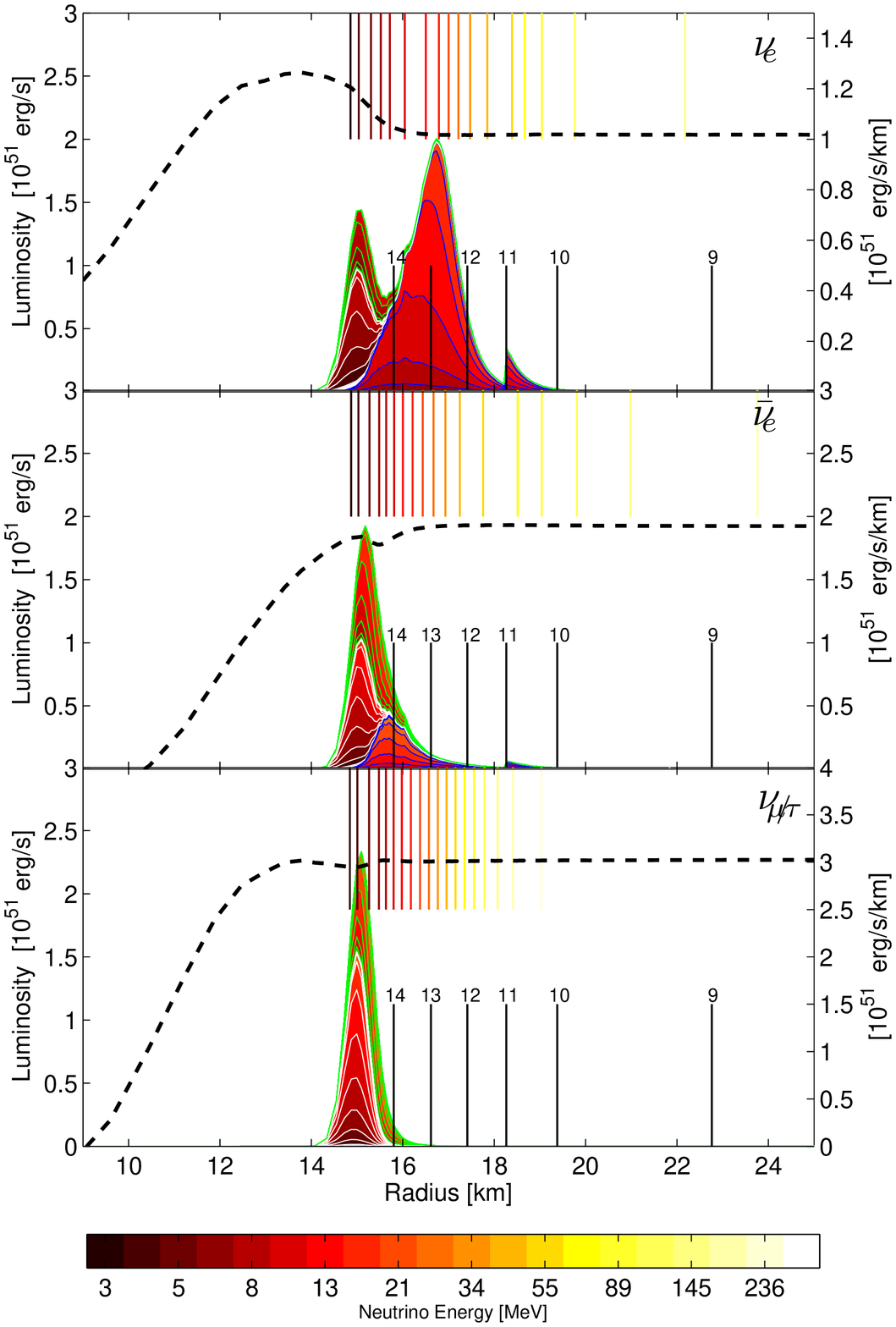}
\label{fig:nustatistics-h18b-a}}
\hfill
\subfigure[\ 22~seconds post bounce]{
\includegraphics[width=0.48\textwidth]{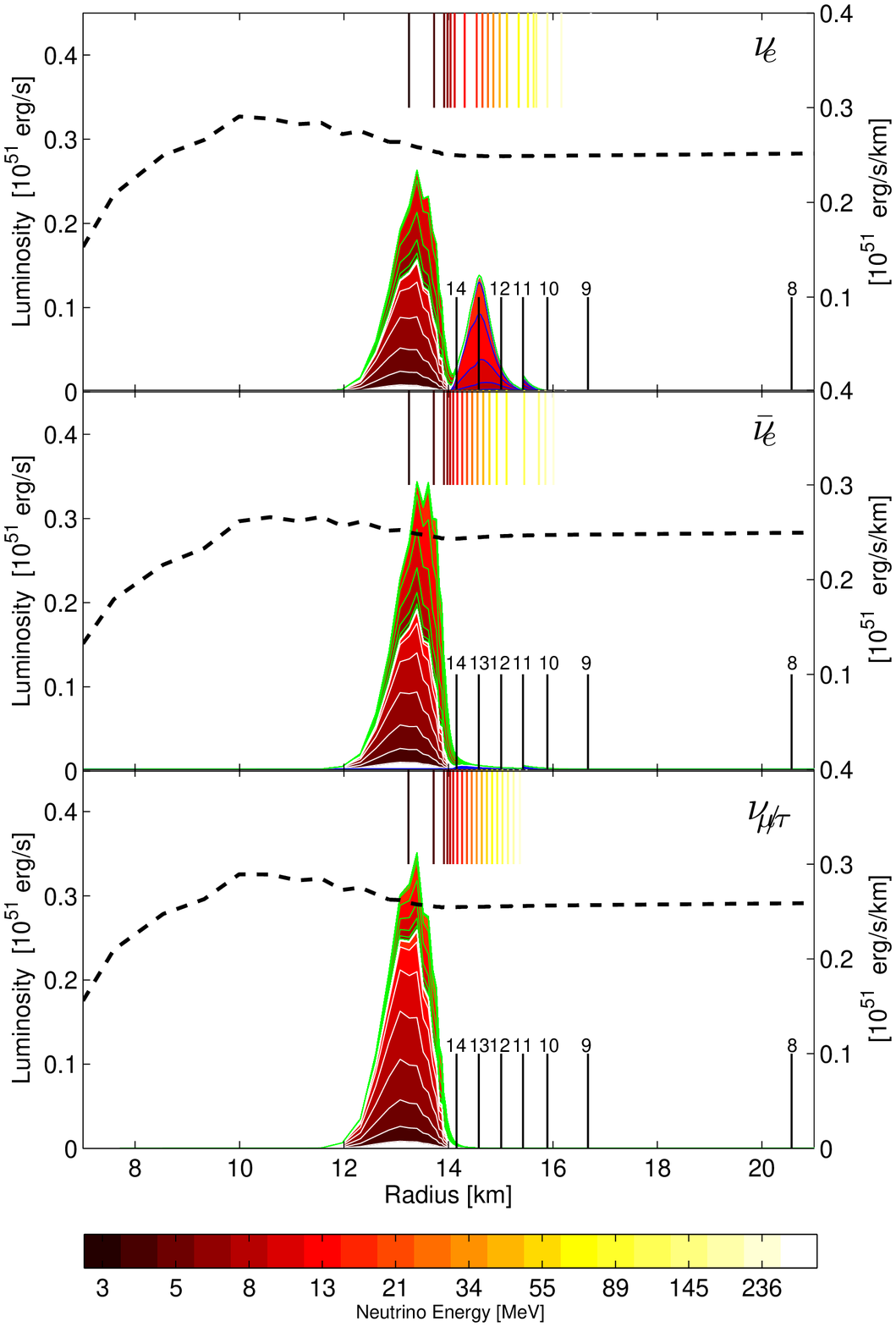}
\label{fig:nustatistics-h18b-b}}
\caption{Local contributions, $\mathcal{L}_i(E,r)$, to the luminosity at infinity (see text for its definition) at two different post-bounce times during the protoneutron star deleptonization for the 18~M$_\odot$ model. The same configuration as Fig.~\ref{fig:nustatistics-n08c}.}
\label{fig:nustatistics-h18b}
\end{figure*}
%

\section{Neutrino emission characteristics}

In the previous sections we analyzed the neutrino spectra based on the opacities, i.e. the propability for a neutrino to be absorbed on its way out of the proto-neutron star. Here we focus on neutrino emission processes, i.e. the inverse of absorption processes, and their contribution to the luminosities at infinity. We follow the formalism of ref.~\cite{Liebendoerfer:2004} where the authors analyzed the early post-bounce phase up to 500~ms. According to appendix~B of ref.~\cite{Liebendoerfer:2004}, we define the quantity $\mathcal{L}_i(E,r)$ such that the luminosity at infinity is given by
\begin{equation}
L_\infty = \int_0^\infty dE \int_0^\infty dr \sum_i \mathcal{L}_i(E,r),
\end{equation}
where $i$ runs over all processes that create a neutrino of energy $E$ at radius $r$. The quantity $\mathcal{L}_i(E,r)$ contains information about the local production of a neutrino of energy $E$ and its absorption during the transport  $r\longrightarrow\infty$.

The function $\mathcal{L}_i(E,r)$ for the different processes considered is shown in the Figs.~\ref{fig:nustatistics-n08c} (8.8~M$_\odot$ model) and \ref{fig:nustatistics-h18b} (18~M$_\odot$ model), at selected post-bounce times. The colors of the contour lines represent the different processes, including charge current (blue), scattering on $e^\pm$ (green) and pair emission (white). Charge-current processes are the inverse of reactions 1-3 of table~\ref{table-nu-reactions} and pair processes are the inverse of reactions 7 and 8 of table~\ref{table-nu-reactions}. Note that elastic processes are not considered as they do not change the energy. The color gradient reflects the neutrino energies, which range from 3 to 300~MeV. The vertical lines at the top of each panel mark positions of the energy spheres and the dashed black lines are the luminosity profiles (left-axis scale). Furthermore, the vertical black lines at the bottom of each panel mark positions of different densities in a logarithmic scale.

As was discussed in ref.~\cite{Liebendoerfer:2004} (see Figs.~(4)--(7)), charge-current processes dominate the emission of $\nu_e$ and $\bar\nu_e$ during the entire accretion phase before an explosion was launched. Here we analyze the later evolution during the proto-neutron star deleptonization.

For the O-Ne-Mg-core illustrated in Fig.~\ref{fig:nustatistics-n08c}, $\nu_{\mu/\tau}$ are produced by pair processes and scattering on $e^\pm$. At early times shown Fig.~\ref{fig:nustatistics-n08c-a}, $\nu_{\mu/\tau}$ are produced by pair processes and their energy is modified by scattering on $e^\pm$. Furthermore, pair processes occur at high densities and hence contribute to neutrinos of higher energies. The produced high-energy neutrinos are down-scattered on $e^\pm$ at lower energies at low densities. The situation for pair processes and scattering on $e^\pm$ is similar for $\bar\nu_e$, however the dominating emission contribution to the luminosity at infinity comes from positron captures on neutrons. It shifts the main production site to lower densities (larger radii) than for $\nu_{\mu/\tau}$. For $\nu_e$, the emission is completely dominated by electron captures on protons at even lower densities. The different production sites reflect the hierarchy of the average neutrino energies.

At later times, illustrated in Fig.~\ref{fig:nustatistics-n08c-b}, we see that charge-current processes for $\bar\nu_e$ become rather small. Hence, $\bar\nu_e$ and $\nu_{\mu/\tau}$ are mainly produced via the same neutral-current processes and at similar densities with slightly different contributions from pair processes and scattering on $e^\pm$. It explains the slight difference between the $\bar\nu_e$ and $\nu_{\mu/\tau}$ spectra (see Fig.~\ref{fig:lumin-n08c}). For $\nu_e$, we observe two clearly separated production sites. Pair processes and scattering on $e^\pm$ occur at almost the same densities as for $\bar\nu_e$ and $\nu_{\mu/\tau}$. However, the second contribution occurs at lower density (larger radius) via electron captures on protons. Both contributions have similar magnitude with the latter one producing neutrinos of lower energy.

For the more massive 18~M$_\odot$ model, the situation is similar at 5~seconds post bounce (Fig.~\ref{fig:nustatistics-h18b-a}). Note the additional small contribution from electron captures on heavy nuclei, at densities on the order of $10^{11}$~g~cm$^{-3}$, which are expected to be present at the neutron-star crust. At later times, illustrated in Fig.~\ref{fig:nustatistics-h18b-b}, charge-current contributions for $\bar\nu_e$ are negligible and the spectra for $\bar\nu_e$ and $\nu_{\mu/\tau}$ become increasingly similar with time. For $\nu_e$, we observe that charge-current contributions decrease continuously and will become also negligible if the simulation was carried out further. It explains the slight difference remaining between the average energies of $\nu_e$ and $\bar\nu_e$ (see  Fig.~\ref{fig:lumin-h18b}). These finding are consistent with the previous discussion based on opacities.

It is interesting to note that pair processes, mainly $N$--$N$--Bremsstrahlung, occur at high densities on the order of $10^{14}$~g~cm$^{-3}$, where the treatment of nuclear correlations may affect the neutrino opacities~\cite{Schwenk:2009}. However, all neutrino flavors will be influenced equally without producing any changes in the relative energies between different flavors.

\section{Summary and conclusions}

We have performed a detailed analysis of the different processes that determine the neutrino spectra of all flavors during the deleptonization phase after the onset of supernova explosion. We have explored the 8.8~M$_\odot$ O-Ne-Mg-core and the 18~M$_\odot$ iron-core progenitors from ref.~\cite{Fischer:2009af}, in order to cover a broad range of stellar models. 

Using neutrino opacities (inverse mean free paths) for the different processes, we computed energy and transport spheres for each neutrino flavor. Our results confirm the finding of ref.~\cite{Raffelt:2001,Keil:2003} for $\nu_{\mu/\tau}$, i.e. the presence of a scattering atmosphere between the $\nu_{\mu/\tau}$ energy and transport spheres. In addition we also find a scattering atmosphere for $\bar\nu_e$ already early after the onset of an explosion, which has not been observed before. We find that at early times after the onset of an explosion, due to the contribution from charge-current reactions to the total opacity for $\bar\nu_e$, the energy spheres of $\bar\nu_e$ and $\nu_{\mu/\tau}$ are located at different positions. However, due to Pauli blocking of final state neutrons the $\bar\nu_e$-opacity from charge current-reactions decreases continuously and the location of the energy spheres for $\bar\nu_e$ and $\nu_{\mu/\tau}$ becomes increasingly similar. At late times, the dominating contribution to the total opacity of $\bar\nu_e$ and  $\nu_{\mu/\tau}$ comes from neutral-current reactions which are independent of the neutrino flavor. It explains the increasing similarity of the $\bar\nu_e$ and $\nu_{\mu/\tau}$ spectra.

On the other hand, the total opacity for $\nu_e$ is dominated by charge-current neutrino absorption on neutrons at early times and hence no scattering atmospheres is present yet. As the charged current opacity for $\nu_e$ is larger than for $\bar\nu_e$, due to the much larger abundance of neutrons compared to protons, neutral-current processes become dominating only at later times than for $\bar\nu_e$. The reduction in the charged current opacity for $\nu_e$ is related to Pauli blocking of the final state electrons, which continuously increase their Fermi energy as the neutrinospheres move to higher densities. At the same time, the average neutrino energies decrease. At these late times, a scattering atmosphere also develops for $\nu_e$, which has also not been observed before, and the spectra of $\nu_e$ and $\bar\nu_e$ become increasingly similar.

In addition to the opacities, we have also explored emission processes (inverse opacities) from which we find qualitatively a similar behavior. Furthermore, we find that charge-current and neutral-current processes contribute to the $\nu_e$ luminosity at clearly separated density domains. Charge-current contributions decrease during the deleptonization and originate from low densities around $10^{10}$--$10^{13}$~g~cm$^{-3}$. On the other hand, neutral-current contributions originate at high densities between $10^{13}$ to several times $10^{14}$~g~cm$^{-3}$ (depending on the progenitor and the state of deleptonization).

A clear extension of this investigation is the inclusion of improved neutrino interactions, in particular corrections from weak magnetism and nucleon recoil~\cite{Horowitz:2001xf}, the emission of $(\mu,\tau)$-neutrino pairs via the annihilation of trapped electron neutrino pairs and the reverse process~\cite{Buras:2002wt}. These were included in the study of the O-Ne-Mg-core in ref.~\cite{Huedepohl:2010}, which was in qualitative agreement with our findings. In addition, the treatment of nuclear correlations~\cite{Schwenk:2009} and weak processes with light nuclei~\cite{Arcones:2008,Sumiyoshi:2008} should also be improved. Nevertheless, we expect that none of these additions will change the findings discussed in this paper, i.e. on timescales on the order of tens of seconds charge-current reactions are suppressed due to Pauli blocking of final states and the neutrino spectra are dominated by neutral-current processes. These processes do not distinguish between different neutrino flavors and hence the neutrino spectra of all flavors become very similar at these late times. Moreover, we expect that including the improved phase-space treatment of ref.~\cite{Reddy:1998} in long-term simulations of supernova explosion, will even enhance the suppression of charge-current processes.

We expect that the neutrino-driven ejecta will always be proton rich. This excludes neutrino-driven winds from non-rotating and not magnetic proto-neutron stars as possible site for the production of heavy $r$-process elements ($Z>56$). Furthermore, the role of neutrino-flavor oscillations at late times for neutrino detection and nucleosynthesis becomes negligible in the presence of very similar neutrino spectra of all flavors. However at early times, when the spectra of the different flavors are still different, it may impact nucleosynthesis under proton-rich conditions~\cite{MartinezPinedo:2011,Wu:2011} and neutrino detection.
%

\section*{Acknowledgements}
The authors would like to thank A.~Arcones and K.~Langanke for helpful discussions. T.F is support by the Swiss National Science Foundation under project~no.~PBBSP2-133378. G.M.P. is partly supported by the Deutsche Forschungsgemeinschaft through contract SFB 634, the Helmholtz International Center for FAIR within the framework of the LOEWE program launched by the state of Hesse and the Helmholtz Association through the Nuclear Astrophysics Virtual Institute (VH-VI-417). is supported by the Swiss National Science Foundation (SNF) under project number no. 200020-132816/1. M. L. and M. H. are also grateful for participating in the EuroGENESIS collaborative research program of the ESF and the ENSAR/THEXO project. The authors are additionally supported by CompStar, a research networking program of the European Science Foundation.
%


\end{document}